\documentclass[useAMS,usenatbib]{mn2e}
\usepackage{graphicx}

\title[NIR Surface Photometry of Barred Galaxies]
{Near-Infrared Surface Photometry of a Sample of Barred Galaxies}

\author[Gadotti et al.]
{D. A. Gadotti$^{1,2,3}$\thanks{E-mail: dimitri@mpa-garching.mpg.de},\vspace{0.35 cm}
E. Athanassoula$^{2}$, L. Carrasco$^{4}$, A. Bosma$^{2}$, R. E. de Souza$^{1}$,\\
\hspace{-0.1 cm}{\LARGE{\rm and E. Recillas}$^{4}$}\vspace{0.25 cm}\\
$^{1}$Departamento de Astronomia, Universidade de S\~ao Paulo,
Rua do Mat\~ao 1226, 05508-090, S\~ao Paulo-SP, Brasil\\
$^{2}$Laboratoire d'Astrophysique de Marseille, Observatoire
Astronomique de Marseille Provence, 2 Place Le Verrier,\\ \, 13248 Marseille Cedex 04, France\\
$^{3}$Max-Planck-Institut f\"ur Astrophysik, Karl-Schwarzschild-Str. 1, D-85748
Garching bei M\"unchen, Germany\\
$^{4}$Instituto Nacional de Astrof\'isica, Optica, y Electr\'onica,
Luis Enrique Erro 1, Tonantzintla, C.P. 72840, Puebla, Mexico}

\begin{document}


\pagerange{\pageref{firstpage}--\pageref{lastpage}} \pubyear{2006}

\maketitle

\label{firstpage}

\begin{abstract}
We have obtained deep J and Ks images of a sample of nine
barred galaxies in order to collect a reliable and homogeneous set
of images to which $N$-body simulations of barred galaxies will be
compared. The observations were performed using the new near-infrared camera
available at the 2.1-m
telescope of the Observatorio Astrof\'isico Guillermo Haro (OAGH) in Cananea,
Sonora, Mexico. We present the results of surface photometry techniques
applied to the observed images, as well as to the deprojected
images. These results include radial profiles of surface brightness
(elliptically averaged),
colour, position angle, ellipticity and the b$_4$ Fourier component. In addition,
we present isophotal maps, colour maps, surface brightness profiles along the bar
major and minor axes, characteristic radial scale-lengths
and bar length estimates. We discuss how projection effects can influence these
measurements and the uncertainties introduced by deprojecting galaxy
images. We show that analytical expressions can be used to obtain
reliable estimates of deprojected bar lengths, ellipticities and position angles
directly from the observed images.
These expressions are based on the assumption that the outer parts of the
bar are vertically thin, as shown by theoretical work.
The usefulness of our data in addressing issues on bar formation and evolution
is also discussed. In particular, we present
results showing a steep drop in the ellipticity profile, as expected for
bar formation processes in which the dark matter halo plays a
fundamental role. Furthermore, we show that the location of this drop is a
good indicator of the end of the bar in strongly barred galaxies, as predicted by
numerical models.
\end{abstract}


\section{Introduction}

It is now widely recognised that bars are one of the major drivers of galaxy
evolution and play a crucial role in shaping the present properties of galaxies
[see \citet{kor04} and \citet{ath05a} for recent reviews of
the observational and theoretical aspects]. Comprehensive
studies, both observational and theoretical, are therefore called for
to allow us to understand best these structures. Furthermore, a strong
connection between bars and the dark matter haloes hosting disc galaxies
was recently revealed. \citet{ath03,ath02} showed that a
significant exchange of angular momentum between near-resonant
particles in the disc and in the dark halo leads to stronger bars.
Thus, at least in principle, the observed properties
of real bars may give us information on the ability of the halo and of
its resonances to absorb angular momentum
from the inner disc, and thus, indirectly, some clues on halo
properties, like its mass and velocity distributions. In
\citet[][hereafter AM02]{AM02} the model
with an initially centrally concentrated halo (their model MH) develops a bar which is much
stronger, longer and thinner than the bar in their MD model, that has a less initially
centrally concentrated halo. A Fourier analysis of the face-on density distribution
of model MH shows that the non-axisymmetric components
are very large compared to those of model MD.
Moreover, also the shape of the halo is linked to the evolution of the bar. Numerical
simulations have shown that an initially spherical halo evolves in shape and becomes
prolate-like in the central parts, as the galaxy forms a strong bar,
and rotates with the same pattern speed as the bar
\citep{ath05a,ath05c,colin,ath07}. \citet{cur99}, \citet{gad03} and
\citet{HSA07} suggest that non-axisymmetric
haloes could trigger the bar instability. In particular, \citet{HSA07}
stress that thus the bar formation time-scale could be very short. On the other
hand, the bar growth at later stages of the evolution can be severely
compromised by a non-axisymmetric halo \citep{EZS02,ber06,HSA07}.
The many intricacies in the bar-halo connection
are as yet not fully understood. However, since triaxial haloes are
predicted by cosmology models in which galaxies form through hierarchical merging
\citep[e.g.][]{col96,HNS07}, the observed properties of bars may eventually lead to clues
concerning the very first assembling of galaxies.

Hence, one might devise methods to indirectly derive the physical properties
of dark matter haloes from the observed properties of the bars they host.
The first step in such a study would be, however, to check how well present
bar models describe real bars. If the models are successful, then
they might indeed give us useful estimates of the true
physical properties of real haloes, via comparisons of
observations of barred galaxies to models with known halo properties.
Hence, a detailed comparison between the structural
properties of real galactic bars and those
arising in $N$-body simulations of barred galaxies is necessary. The first challenge
in the pursuit of such a comparison is to convene an appropriate set of galaxy
images. For a reliable comparison with $N$-body models, these images must comply with
certain criteria. Ideally, they should be deep and in the near-infrared (NIR)
wavelength, so that a given galaxy
image is a true representation of the bulk of its stellar population. This is a
fundamental point since the overall evolution of an $N$-body simulation is driven
by the gravitational potential created by the model as a whole. Evidently, such a sample
of galaxy images would also benefit from a homogeneous
treatment, that simplifies the interpretation of the results. In this respect, the 2MASS survey
(see http://www.ipac.caltech.edu/2mass/) is advantageous on a
statistical level.

\citet{men07} measured the bar fraction and the relative sizes of bars
and discs using 2MASS images of 151 nearby spiral galaxies. In another
recent NIR study on bars, based this time on the OSUBSGS,
\citet{mar07} made a similar analysis based on 180 spirals and found that most bars have
moderate to high ellipticity. The main aim of these studies is to provide a
reference for comparisons with galaxies at higher redshifts. Our aim is
totally different; we want to provide a sample for detailed comparison
of observations to $N$-body simulations. We therefore opted for a much
smaller sample, and made a detailed analysis based on individual
study and inspection of each case. The importance of this for our
specific purposes will be made clear during this work. For the same
reasons we aimed at obtaining deeper images.

In this paper, we describe an effort to acquire a suitable set of deep
NIR images of barred galaxies which will be used to study
the morphological properties of bars and of their host
galaxies. With these images,
ellipse fits are possible down to isophote levels $\approx$ 1 mag fainter, on average,
than the detection limit in 2MASS images, both in J and Ks.
To achieve this depth we had to limit our sample size. Furthermore, we
avoid relying on automated procedures. Instead, we do the analysis
for each galaxy individually, examine each case separately and assess
all the results by eye.
This is generally not feasible if the sample contains many galaxies.
Here we present results based mostly on ellipse fits, and postpone to a future paper
the analysis of Fourier components and bar strength \citep[e.g.][]{lau04,but05,but06}.

Our images were obtained with the NIR
camera of the 2.1-m telescope of the Observatorio Astrof\'isico Guillermo Haro (OAGH) in
Cananea, Mexico. The observatory is operated by the Instituto Nacional de Astrof\'isica,
Optica, y Electr\'onica (INAOE) and details about the Cananea Near-Infrared Camera
(CANICA) will be presented in L. Carrasco et al. (2007, in preparation).
The general properties of the galaxies in our sample and the steps taken in the observations
and data reduction are described in Sects. 2 and
3, respectively. Applying surface photometry techniques, we determine many
physical properties of the bars and of the galaxies, which are presented in Sect. 4 and
discussed in detail in Sect. 5. Section 6 deals specifically with bar length measurements.
In Sect. 7 we briefly discuss features and uncertainties in deprojected galaxy images and
the use of these data to assess whether, and up to what extent, results of up-to-date
$N$-body simulations of barred galaxies agree with the photometric
observations. We summarise and conclude in Sect. 8.

\section{The sample}

Our sample consists of 9 galaxies whose relevant properties are given in Table 1. All of them
were observed in J and Ks. Since we aimed for deep NIR imaging, we have not looked
for a complete, unbiased sample, but instead we chose suitable galaxies for our scientific goals,
namely a comparison to $N$-body simulations of barred galaxies.
Our target galaxies
also had to comply with the apparent size limit imposed by the undistorted
field of view of the camera ($\approx 3$ arcmin -- see Sect. 3), since we wanted to avoid
doing time consuming mosaics.
All galaxies are local, most are bright, many relatively close to
face-on and they span
a range in morphologies. According to the RC3 \citep{dev91}, three are SAB (weakly barred) and six are
SB (strongly barred). Six galaxies have morphological types S0 or S0/a
and the remaining three go as late-type as Sb. Table 1 also shows that four galaxies in our sample
have nuclei with non-stellar activity (AGN).

The choice of local, bright and moderately inclined galaxies gives us more reliable estimates
for the structural parameters of these galaxies, since it means higher signal-to-noise ratio
and higher spatial resolution, while in more edge-on systems a proper description of some bar
properties may be unattainable. Because deprojected measurements are preferred for a suitable
comparison of real images to $N$-body realizations, we avoided highly inclined systems.
This assures us that spurious geometric effects introduced by image
deprojection techniques are avoided for most of our galaxies.

The diversity of our sample might be helpful in trying to evaluate
clues related, for instance, to the prominence of the classical bulge and the bar strength.
The presence of galaxies with AGN might also be relevant
to help in understanding the role played by bars in the fuelling of such nuclei. We also
note that, except for NGC 799, no galaxy in our sample is
part of a pair, or of a multiple system,
which means that their bars are most likely fully due to internal instabilities,
with no significant help from tidal forces.
This assessment was made using LEDA (the Lyon Extragalactic Data Archive)
and publicly available images in the NASA Extragalactic Database (NED), but
does not exclude the possibility that these galaxies are members of galaxy clusters.
NGC 799 has an interacting companion similar in size and luminosity at a projected distance
of the order of its own diameter. There are, however, no signs of
a violent interaction, which suggests it might be in an early stage,
or that the passage may be retrograde, or the deprojected separation
much larger than the projected one.

\section{Observations and Data Reduction}

The NIR images of the galaxies in our sample were obtained with CANICA, a camera
based on a HAWAII 1024 x 1024 pixel array, available at the 2.1-m telescope of the OAGH,
in Mexico. The plate scale and field of view are 0.32 arcsec per pixel
and about 5.5 x 5.5 arcmin, respectively (note, however, that the
outskirts of the field of view suffer from optical distortions,
meaning that, for our purposes, the safest procedure is to avoid galaxies larger than approximately
3 arcmin in diameter). An electronic cross-talk effect is present in CANICA at a
0.9\% level, which might cause photometric errors if there are bright sources in a given field.
This means that, if there is a bright star in the field of the galaxy, a very small fraction
of the star light can be spread through a few consecutive lines in the array and, in some cases,
cross the galaxy image. Unfortunately, this happened to the images of
two of our galaxies. Although the effect
may be negligible, we took all the necessary steps so that our
results are not affected and we present the corresponding procedures at the end of this section.

The data were obtained during a single run of
12 nights in September and October 2004, when all galaxies were observed in both J and Ks.
Table 2 shows relevant data on the observing run. Except for
4 nights possibly with small clouds, most of the nights were clear and photometric.
This made our photometric measurements quite accurate. Every night, several
standard stars from \citet{hun98} were observed in both bands. The photometric error
for each night was assumed to be the standard deviation between our estimated magnitude
and the magnitude determined in \citet{hun98} for the standard stars observed. This means that,
since during some nights there were more stars observed than in others, the error estimates
have different accuracies. This may explain why the error in the photometric nights
is typically not too different from that in the non-photometric
nights. The mean zero-point error in J in the photometric nights is 0.06 mag,
reaching 0.08 mag in the non-photometric nights. Similarly, in Ks we
have errors of 0.08 mag and 0.10 mag, respectively. Coincidentally,
in the non-photometric nights we had a full moon increasing the sky brightness, but that in fact
does not seem to harm our measurements, as expected for NIR observations.
We lost, however, some images due to the direct incidence of moon light in the dome.
As expected, the error estimates show that the photometric accuracy
is generally higher in J than in Ks. We note that our average errors in the photometric
zero-point are 0.07 mag in J and 0.09 mag in Ks, which is quite good for NIR bands.

Our observations, nevertheless, have one drawback, namely
the seeing, which was poor during the whole run. The FWHM of Gaussian
fits to the light profile of standard stars observed in the J-band was constantly
around 2.0 arcsec. Moreover, after co-adding all images taken for a given galaxy in a given band
the spatial resolution got poorer by typically 0.5 arcsec, due to uncertainties in the process
of combining many images. All galaxies were observed during two or more nights
to reach the total integration time aimed for, and also sometimes to replace images that
were found to have problems, such as due to a bad telescope move or
when the galaxy image was out of the undistorted field of view of the camera.

Using results from previous NIR observations \citep{gad03,gad06}, we designed an
observing strategy in order to reach a signal-to-noise (S/N) ratio of $\sim3$ at the 21 Ks
mag arcsec$^{-2}$ isophote, although we could not be very precise since technical parameters
like the camera efficiency were not known, because the data obtained as part of its
commissioning were not fully reduced and analysed. Currently, CANICA's technical details
can be found in the OAGH web pages. Nevertheless, the ellipse fits to the isophotes of
our images described further below reach on average $20.1\pm1.1$ Ks mag arcsec$^{-2}$
and $21.4\pm1.2$ J mag arcsec$^{-2}$. To reach this S/N we needed a total exposure
time on target of 6000s per galaxy per band, which was in fact achieved. In a few cases
the total integration time is a bit smaller due to problems in some images.

Since NIR images are limited by background emission, we used the following observing
strategy. We define as a cycle a set of 18 exposures of 50s each, starting with two consecutive
exposures on target, followed by two on sky and so on. In Ks we ran 12 such cycles per galaxy.
The interwoven sky exposures are necessary for a proper background
subtraction, since the background intensity
can change significantly in time-scales as short as a few minutes. As it can also
change rapidly spatially, the sky images were taken only a few arcminutes from the galaxy.
The sky images are dithered a few arcseconds in order to ease the removal of unwanted objects.
Similarly, we dithered the galaxy images in order to avoid the effects of bad pixels.
Since the sky background is fainter and more stable in J, the individual exposures in this
band reach 150s, so that only 4 cycles per galaxy were needed. For the standard stars, the background
contribution can be well estimated within the star image frame and typically we made 8
dithered images of around 10s each to get the final star image.

Before actually going through all data reduction steps {\em every}
image was checked for problems, even though the number of image files is very large
(72 per galaxy in J, 216 per galaxy in Ks).
For the treatment of the images we used the {\sc gemini} {\sc iraf}\footnote{{\sc iraf}
is distributed by the National Optical Astronomy Observatories,
which are operated by the Association of Universities for Research in Astronomy, Inc., under
cooperative agreement with the National Science Foundation.} package. Flatfield images
were obtained through the {\sc qflat} task from combining many dome images. We decided to use
dome flatfields, instead of flat images obtained from sky images, after checking with many
standard stars that the former produced more accurate results in terms of photometry
(i.e., smaller zero-point errors). The {\sc qsky} task was used to estimate the background
contribution. For each galaxy image, the background was estimated from the 4 sky images closest
in time. Furthermore, the mean and standard deviation of every sky image were calculated, and
if the mean was discrepant from that of the other 3 sky images by more than 10\% of
the standard deviation then the image was removed from the process. In this way, we
avoid background changes and ensure a proper background subtraction. Interestingly
enough, the standard deviation of the sky images was very similar even when
there were (small) changes in the mean, and typically only a few background images
had to be removed per galaxy. These corrections (flat-fielding and background subtraction)
were performed for every galaxy and standard star image by {\sc qreduce}, using the appropriate
flatfield and background images.
Finally, the {\sc imcoadd} task combines by the median all corrected images
of a galaxy, calculating the necessary shifts due to the dither pattern.

In an effort to avoid spurious effects caused by the electronic cross-talk mentioned above, the J
and Ks images of NGC 1358 and NGC 7743 (in this study, the only galaxies whose images are
affected by this problem) underwent further processing. This consisted of the following steps.
First, the lines of the detector having the problem were identified (by visually inspecting
the images) and masked out. Next, using the {\sc ellipse} and {\sc bmodel} tasks in {\sc iraf}
a simple model image was fitted to the galaxy. The ellipse fits were performed in a similar way as
those presented below. Finally, the bad lines were substituted by the corresponding lines
from the model.

\section{Surface Photometry Analysis}

\subsection{General Presentation}

Figure \ref{gals} shows for each galaxy the J-band direct image, as well as
J$-$Ks colour maps with J-band isophotal
contours overlaid (with a difference of 0.5 mag between two consecutive contours).
The latter were built by dividing the J image by the Ks image
(both in ADN units) after trimming and alignment. Since the PSF FWHM is very similar
in both bands, there was no need for degrading the PSF of any of the images.
The grey scales in these figures vary from galaxy to galaxy
since they were chosen to emphasise each galaxy's main features. These
figures show clearly the deepness of our images,
witnessed for instance by the fact that one is able to trace the spiral arms (e.g. in NGC 266) for quite a
large azimuthal angle, which is not usual in the NIR. A similar evaluation can be done from the data
in Table 1 and Figs. \ref{res266} to \ref{res7743} below,
as well as from a comparison with images from 2MASS in NED.

Figure \ref{barprof} shows the surface brightness profiles along the major and minor axes of the bars of the
galaxies in our sample. Most of the profiles along the bar major axis show the typical shoulders
found in optical images for bars (especially strong ones) in galaxies with morphological
types earlier than Sbc \citep{elm85}.

We fitted ellipses to the isophotes of each galaxy in both bands,
using the {\sc iraf} task {\sc ellipse}.
The increase in the semi-major axis between two consecutive
isophote fits is 1 pixel (0.32 arcsec). We thus built radial profiles
of the elliptically averaged surface brightness, and of
geometric parameters of the isophotes, namely, position angle,
ellipticity and the b$_4$ Fourier coefficient. These geometric
parameters are practically identical in the two bands so we just show (Figs. \ref{res266} to \ref{res7743}) those
relative to the J band, where the S/N is better. In addition, we present J$-$Ks colour profiles.

\subsection{Ellipse Fitting}

During the ellipse fitting procedure the centre was held fixed. To find the galaxy centre, we first ran
{\sc ellipse} with the centre free and then chose as centre a typical value from those given by
the task for the isophotes at a radius around 10 to 15 pixels. We found that the dispersion of
the values of the central coordinates obtained from this region is generally lower
(i.e., only $1-2$ pix) than the corresponding dispersion at larger radii.
In addition, this region is far enough from the centre to make sure that
poor statistics do not result in an ill-defined centre. After fixing the centre, the error bars given
by {\sc ellipse} for the coordinates of the centre are of the order of one pixel.
Similar results are obtained for both the $x$ and $y$ coordinates.
Interestingly, the location of the galaxy centre determined in this way is often
identical to that of its brightest pixel (and when it is not, the differences are below $1-2$ pixels).
Similar results are obtained using the {\sc imcentroid} task in {\sc iraf}.

One may ask whether
different results could arise from the ellipse fits if the centre was left free.
To check that, we inspected the ellipse fits to the J-band image of NGC 266, when letting the
central coordinates vary. It turned out that the results do not change significantly
over most of the galaxy. The position of the centre of each isophote
varies only by a few pixels, which does not cause substantial changes to the other
relevant radial profiles. When one reaches the outer spiral arms, however,
the central coordinates might assume completely wrong values, by as much as several tens
of pixels, and thus will of course affect all radial profiles
in this region. This is due to the asymmetric nature
of the arms. We thus always held the centre fixed in the ellipse fits presented here.

The ellipticity of an isophote is defined as $1-b/a$, where $a$ and
$b$ are, respectively, the semi-major and semi-minor axes of the
best fitting ellipse. The b$_4$ coefficient is related to the fourth harmonic term of the Fourier
series that fits the deviations of the isophote from a pure ellipse \citep[see also the {\sc iraf}
help pages]{jed87}. It is the amplitude that multiplies the term $\cos(4\theta)$ normalised
by the isophote semi-major axis length and the local intensity gradient (where $\theta$ is
the azimuthal angle). This coefficient thus measures deviations from a pure
ellipse that are due to either boxy (b$_4<0$) or discy (b$_4>0$)
isophotes. Edge-on bars produce
boxy isophotes \citep[e.g.][and references therein]{des87,kui95,burf99,bur05}.
In face-on barred galaxies the superposition of the bar and
a prominent classical bulge may create discy isophotes \citep[see][and the results
below]{des04,gad06,ath90}. If, however, the bulge component is carefully
masked out, then, at least for strongly barred early-type galaxies,
the isophotes have a strong rectangular-like shape \citep{ath90}.
This argues that the intrinsic shape of bar isophotes is in fact
rectangular-like, while the discy outlines are due to the
superposition of the bulge component. This argument is further
strengthened by ellipse fits to $N$-body bars (AM02), where the
classical bulge component can be easily removed and fits can be made to the disc only,
or to the disc plus bulge components.

Similarly, the a$_4$ coefficient is the amplitude of the $\sin(4\theta)$ term.
It is important to stress that to measure the strength of the $m=4$ Fourier component
in the galaxy image one has to account for the contribution of both
terms. This is always done when a$_4$ and b$_4$ are measured using
circular concentric rings \citep[see e.g.][]{oht90,AM02,but06,lau06}, but is
neglected when they are measured from ellipse fitting. Indeed, in the
latter case it is implicitly assumed that the $m=2$ and $m=4$
components have roughly the same position angle. If this is true,
one expects a$_4$ to be negligible, since the fitted ellipses will
be aligned with the bar in the bar region. We checked this assumption
for our sample and found that it holds for 5 galaxies: in the bar region, the maximum of a$_4$
is much smaller than that of b$_4$ (smaller by a factor of 5, or more). However,
for NGC 266, 357, 7080 and NGC 7743, the ratio of the a$_4$ peak to the b$_4$ peak is,
respectively, 0.04/0.06, 0.04/0.12, 0.03/0.14, and 0.02/0.04. Hence, the contribution
from a$_4$ can be as large as half that of b$_4$, even when these parameters are estimated through
ellipse fits, rather than using circular concentric rings.
Interestingly, we note that all galaxies with a significant contribution
from a$_4$ have conspicuous spiral arms or rings, which might contribute to the
$m=2$ and 4 differently than the bar. These results show that it can be hazardous to
straightforwardly neglect the contribution of the a$_4$ component.

\subsection{Comments on Individual Galaxies}

Our images of NGC 357, NGC 1211 and NGC 7280 reach the 25 B-band mag arcsec$^{-2}$ isophote. On the other
hand, in the cases of NGC 799 and NGC 1358 our surface photometry analysis goes approximately
only half as far. For the former, an SDSS image (Sloan
Digital Sky Survey -- http://www.sdss.org) shows that the bar is
relatively weak and is embedded in a lens.
It is interesting to note that \citet{erw03} found a secondary bar in NGC 7280,
which is likely the cause of the first peak we find in its ellipticity profile in the nuclear region
(at $\approx$ 1 arcsec from the centre -- Fig. \ref{res7280}).
In addition, \citet{erw02} comment that NGC 7743 has a
nuclear spiral structure which stands out clearly as a blue feature in the colour map and colour
profile we present here (Figs. \ref{gals} and \ref{res7743}, respectively).
In this case too, this might be the cause of
the ellipticity peak we find in the nuclear region ($\approx1-2$ arcsec). Note, however, that seeing
effects are important at these distances from the centre in all our images.

The isophotal contours of NGC 1638 display a considerable asymmetry which
seems to be real, and not an artifact from wrong sky
subtraction, since it is also present in the images available in
NED. The origin of this asymmetry might be related to the presence of
dusty spirals over the northeast corner of the galaxy.

In the images and colour maps of NGC 1358 and NGC 7743 it is possible to identify the detector lines
corrected by the electronic leak problem discussed previously, and whose correcting
procedure is presented in Sect. 3. The changes introduced by these corrections in the
original galaxy image are not in terms of brightness or brightness
gradient, but in the absence of noise. This was verified by analysing intensity cuts
parallel and perpendicular to the corrected lines, as well as along them. This is due to the
fact that these lines were generated by a model which does not include noise. Since the brightness
range in the colour maps is very narrow (only a few tenths of magnitude)
this difference in the corrected lines stands out more clearly in these maps
than in the direct images. Nevertheless, it is important to stress
that, since the correction was generated by models from ellipse fitting, this problem
does not introduce spurious effects to the results presented in this paper,
which are themselves based on ellipse fits to the galaxy images.

\subsection{Deprojecting the Images}

The surface photometry techniques applied to the observed images were also applied
to deprojected images of all galaxies in our sample (results shown in red in Figs. \ref{barprof} to \ref{res7743}).
To deproject each galaxy image we performed a flux conserving stretching of the direct images
in the direction perpendicular to the line of nodes, using the {\sc iraf} task
{\sc imlintran}. As position angle of the line of nodes (PA$_{\rm ln}$), we adopted that of the
25 mag arcsec$^{-2}$ isophote in the B-band from RC3, except for NGC 266, 357, 1358 and
NGC 7080. For these galaxies this information is not present in the RC3, and so we considered
the position angle of our faintest isophote fit.
The inclination angles $i$ were taken from LEDA, except for NGC 7743 (see below).
Since errors in the estimates of PA$_{\rm ln}$ and $i$ can
lead to wrong results when deprojecting galaxy images, it is important to check whether
our values agree with those of other sources, when available. For PA$_{\rm ln}$, we checked estimates
from LEDA and visual inspection of deep blue and red images from the
Second Generation Digitized Sky Survey (DSS2), available, e.g. at {\sc skyview}
(http://skyview.gsfc.nasa.gov). For NGC 799 and NGC 1211 we also inspected images from the SDSS.
All estimates agree within $\approx10^\circ$
except for NGC 266, NGC 1358 and NGC 7080. For NGC 1358, the
position angle from LEDA is 49$^\circ$, while the estimates both from
our images and from the DSS2 images point to 15$^\circ$. We stick to our estimates and suggest that
the LEDA value results from shallower images, as one can conclude from Fig. \ref{res1358}, where one sees
that 49$^\circ$ is the position angle of the isophotes at about 27 arcsec, going to 15$^\circ$
further away from the centre. Likewise, the PA$_{\rm ln}$ values of NGC 266 and NGC 7080 quoted in LEDA
(95$^\circ$ and 100$^\circ$, respectively) are significantly different from the estimates
both from our images and from the DSS2 images.
These values, however, were determined from the shallower images of the 2MASS survey and clearly
refer to an inner part of the galaxies, as can be checked through visual inspection of the images
and from the results of our ellipse fits in Figs. \ref{res266} and \ref{res7080}.

It is also possible to use the DSS2 and SDSS images to measure the axial ratio
of the outer isophotes and to
obtain another estimate for $i$, assuming that the outer disc should be intrinsically circular.
For most of our galaxies these estimates agree with those
from LEDA, with a difference of less than about 20$^\circ$. The two exceptions are NGC 266,
which appears to be more inclined in the DSS2 images,
and NGC 7743, which appears to be less inclined. For NGC 266,
since \citet{ma98} give an angle very similar to LEDA (12.2$^\circ$) we adopted the LEDA estimate.
For NGC 7743, we obtained from the DSS2 images an axial ratio $b/a=0.85$, resulting in an
inclination angle of $\approx31^\circ$. This is a factor of 2 lower than the value quoted in LEDA.
We decided to discard the LEDA value and use our estimate for the inclination angle of this galaxy since
it agrees with the value quoted in \citet{erw05}. We found in the literature a different source
for the inclination angle of two other galaxies in our sample. For NGC 357, \citet{erw04} gives
a somewhat lower inclination (37$^\circ$), but still within 10$^\circ$ from the LEDA value.
For NGC 7280, \citet{erw05} gives an angle of 48$^\circ$, again similar to the LEDA value.
Table 1 lists our adopted values for PA$_{\rm ln}$ and $i$.

\section{Radial Profiles}

\subsection{Surface Brightness Radial Profiles}

In this paper we present three different types of photometric radial profiles.
In Fig. \ref{barprof} we show those along the major and minor axes of the bars. They were built by
extracting intensity counts from the images along two narrow strips, each over one
of the bar axes. The position angle of the bar major axis, PA$_{\rm bar}$, was estimated from the
ellipse fits to the isophotes in the bar region (see Figs. \ref{res266} to \ref{res7743}). The strips have a
width of five pixels, over which an average is calculated to obtain the final intensity value
at a given galactocentric distance. Evidently, the contribution of the bar is maximised
in the luminosity profile along its major axis, making this type of profile especially
suitable to study bars. The disc contribution is only clearly seen after the end of the
bar, especially in the case of a strong bar. This means that to make a better
assessment of the disc component in such a case one has to look at the
luminosity profile along the bar minor axis, where the contribution of the bar is minimised.

In Figs. \ref{res266} to \ref{res7743} we show surface brightness radial profiles obtained from the ellipse
fits. The intensity at each point is an azimuthal average over the
fitted ellipse and its galactocentric distance is the length of the
ellipse semi-major axis. Note that the ellipticities and position angles of the fitted ellipses vary.
This means, for instance, that these luminosity profiles are {\em not} calculated over
a straight path from the centre to the outskirts of the galaxy. Thus, although the contributions
from each individual galaxy component are added together in this type of luminosity
profile, such a profile is still very useful, since it
depicts the major component in the different regions of a galaxy. For barred disc
galaxies like those studied here, it is clear that while bulge and disc dominate the inner and
outer parts, respectively, bars can be the dominant component at intermediate distances from
the centre.

Near-infrared imaging is not optimum for studying the outermost part of discs. However,
inspecting the surface brightness profiles along the bar minor axis in Fig. \ref{barprof}
one notices that for all galaxies in our sample the disc can be suitably described with
a pure exponential law \citep{fre70} until the limits of our measurements. The bulge
contribution in the inner part is also clearly seen in all galaxies, regardless of
the type of luminosity profile.

Although the drops associated with the end of the bar are easier to detect in the
luminosity profiles along the bar major axis in Fig. \ref{barprof}, they are also present
in many of the elliptically averaged profiles in Figs. \ref{res266} to \ref{res7743}
(see NGC 357, 1211, 1358 and NGC 7080).
This suggests the use of these drops to measure bar lengths, which will be discussed
further below, since these drops are due to the smaller amount of light
coming from outside the bar region. It is interesting to note
how all luminosity profiles change in the deprojected images (especially, of course, for the
more inclined galaxies). In particular, when the bar position angle is not close to the
position angle of the line of nodes, bars can get longer when deprojected.
Thus, the location of the drops associated with the end of the bar can change when the images
are deprojected. Interestingly, the slope of the profile after these drops can get less steep as
a result of the image stretching.

\subsection{Ellipticity Radial Profiles}
\label{sec:ellipticity}

In almost all cases, the ellipticity profiles display a clear
systematic behaviour and follow a well defined pattern. Let us
consider first those from the observed (projected) images. After the nuclear
region, for which we can not draw any firm conclusion due to insufficient
resolution, the ellipticity increases steadily, often quasi-linearly,
from $0.1-0.2$ to a high value ($0.4-0.6$), and then stays nearly
constant forming a plateau. In NGC 7743 the extent of this plateau is very
small, but in other cases it is considerable.
For example, in NGC 1211 it is around 8 arcsec (1.7 Kpc). The extent of this
region is largest in NGC 266, where it is 23 arcsec (7.6 Kpc).
After this plateau there is a steep drop, as one would expect from a
very sharp transition between an elongated component (e.g. a bar) and
a near-circular component (e.g. a ring or a disc).
In some of our galaxies, like NGC 266 and NGC 7080, this drop is so
steep that we have no isophotal fits in the corresponding very narrow
radial extent. After this steep drop the ellipticity increases again,
but to a smaller value, which is a function of the inclination and
intrinsic ellipticity of the disc. Three galaxies clearly deviate from this
pattern: NGC 799, NGC 1638 and NGC 7280. They will be discussed further below.

A similar clearly defined pattern is seen in the ellipticity profiles
of the simulations of AM02 (we refer the reader particularly to their Fig. 4). MH-type
models display a plateau in the ellipticity values followed by a very sharp
drop, as observed in most of our galaxies. AM02 used the location of
this drop as one of the possible ways to measure the end of their
MH-type bars. On the other hand, in MD-type models the decrease in
ellipticity with radius is much more gradual.

Two of our galaxies, NGC 266 and NGC 7080 (and to a lesser extent also
NGC 1211), have a second plateau at radii larger than the first one.
It is interesting to note that these galaxies are those exhibiting the
sharpest drops after the first plateau.
A careful analysis of the images and the corresponding isophotal
ellipse fits suggests that these plateaus are caused by a second
component, also non-axisymmetric but not as much as the
bar. We hesitate to call these components a lens, because their extent
is longer than that of the bar, contrary to what was found by \citet{kor79}
for lenses. We will, lacking a better term, call them oval
discs. They are just outside the bar, where the
stellar orbits are expected to be less eccentric \citep{ath92,pat03} and
they have a somewhat different position angle than that of the bar
(except in NGC 1211). A B-band image of NGC 266 shows no clear lens, and argues that this oval
disc is surrounded by the inner spiral arms, while the optical SDSS
image of NGC 1211 shows an inner ring/lens at a position similar to
that of the second plateau, a feature absent in our NIR images.
The second ellipticity plateau in NGC 7080 may be influenced by the
spiral arms, since the position angle of this second structure
varies somewhat with radius in a smooth way, as is generally expected
for spiral arms. Note, however, that the position angle of the
isophotes that describe its spiral arms and outer disc changes much faster.
In these three galaxies there is a second steep drop in ellipticity after the oval disc
(this drop is less sharp in NGC 7080). After the second drop, the ellipticity follows the
general trend found in the outer parts of the other galaxies, increasing steadily to a small value
that reflects the disc inclination and intrinsic ellipticity.

The deprojected ellipticity radial profiles show a very similar pattern, although,
as expected, the ellipticity values and the position of the plateau/peak
might change. NGC 357 is a nice example of these
changes: the bar becomes longer and more eccentric when deprojected, whereas the
outer disc becomes more circular, as expected (see Fig. \ref{res357}).
A noticeable exception is NGC 1211, which shows more eccentric outer isophotes in the
deprojected image. There is a clear explanation, however:
the fit does not reach
the outer, rounder disc present in the direct image, but rather goes up only to
the second ellipticity plateau, caused by the oval disc (see Fig. \ref{res1211}).

\begin{figure*}
   \centering
   \includegraphics[keepaspectratio=true,clip=true,width=4cm]{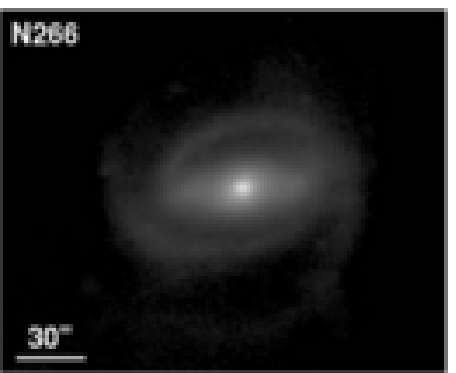}
\includegraphics[keepaspectratio=true,clip=true,width=4cm]{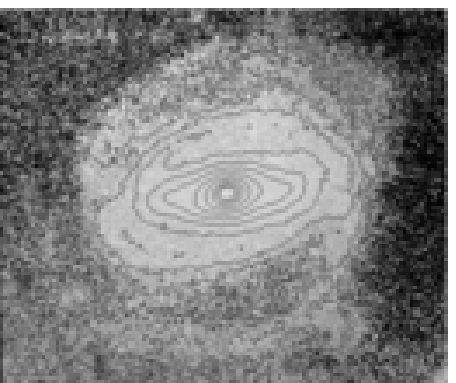}\hskip 1cm
   \includegraphics[keepaspectratio=true,clip=true,width=4cm]{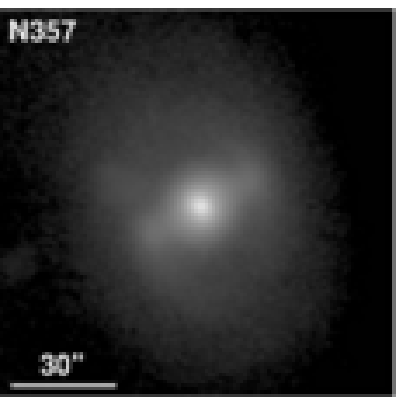}
\includegraphics[keepaspectratio=true,clip=true,width=4cm]{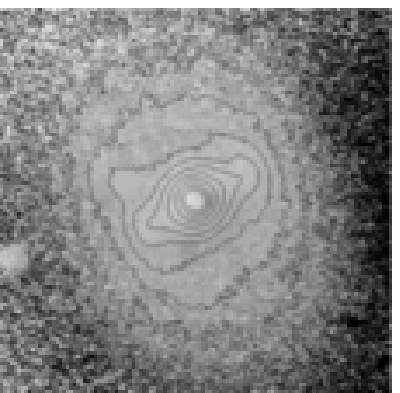}\\
   \includegraphics[keepaspectratio=true,clip=true,width=4cm]{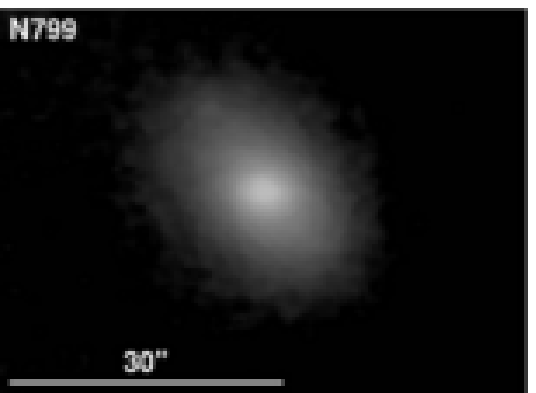}
\includegraphics[keepaspectratio=true,clip=true,width=4cm]{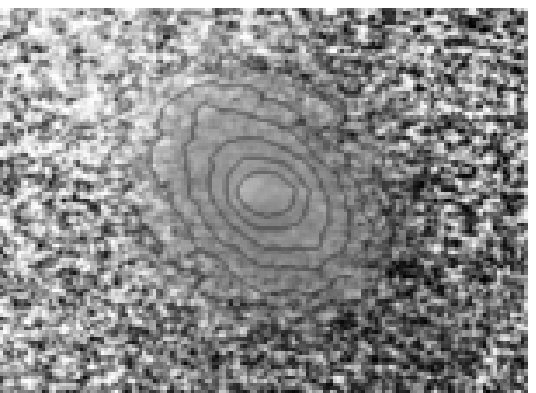}\hskip 1cm
   \includegraphics[keepaspectratio=true,clip=true,width=4cm]{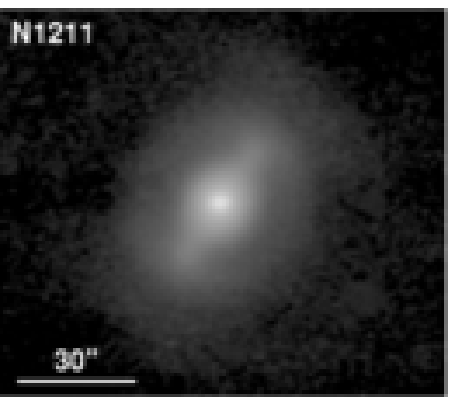}
\includegraphics[keepaspectratio=true,clip=true,width=4cm]{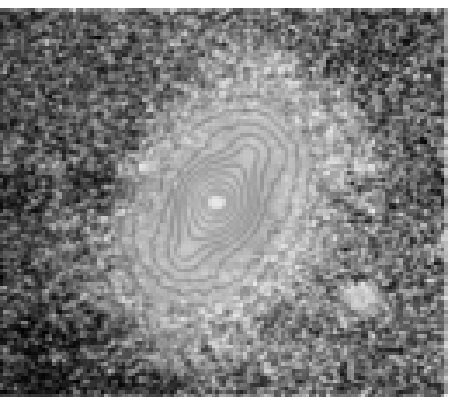}\\
   \includegraphics[keepaspectratio=true,clip=true,width=4cm]{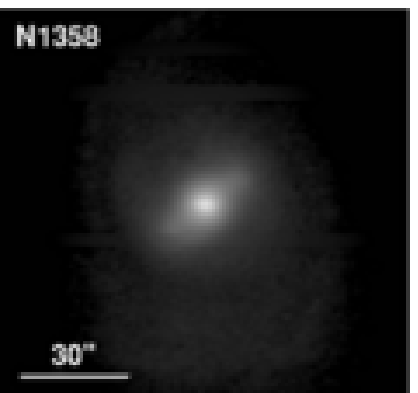}
\includegraphics[keepaspectratio=true,clip=true,width=4cm]{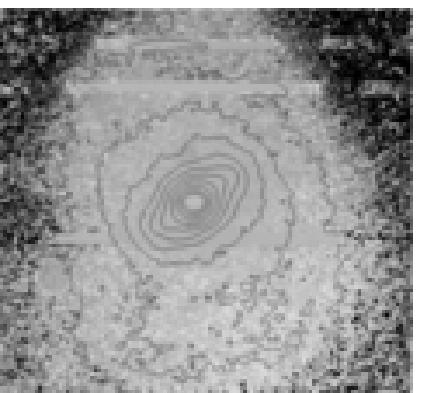}\hskip 1cm
   \includegraphics[keepaspectratio=true,clip=true,width=4cm]{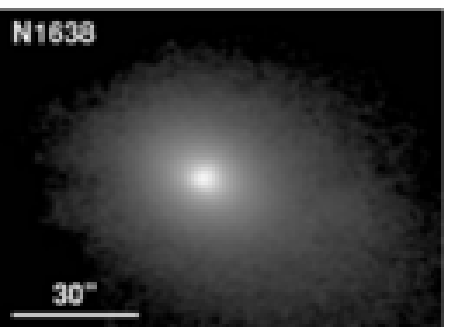}
\includegraphics[keepaspectratio=true,clip=true,width=4cm]{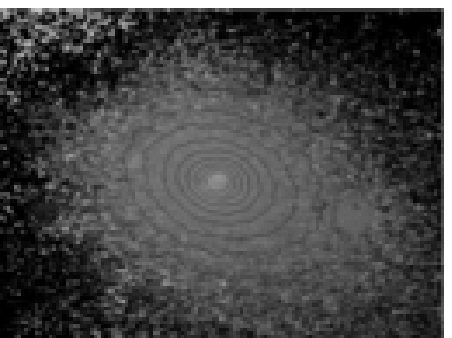}\\
   \includegraphics[keepaspectratio=true,clip=true,width=4cm]{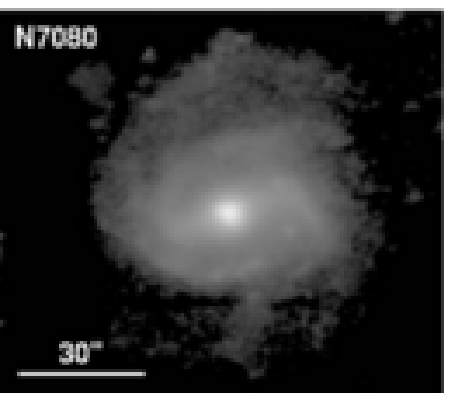}
\includegraphics[keepaspectratio=true,clip=true,width=4cm]{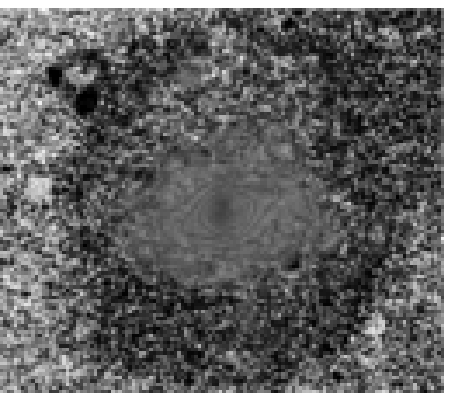}\hskip 1cm
   \includegraphics[keepaspectratio=true,clip=true,width=4cm]{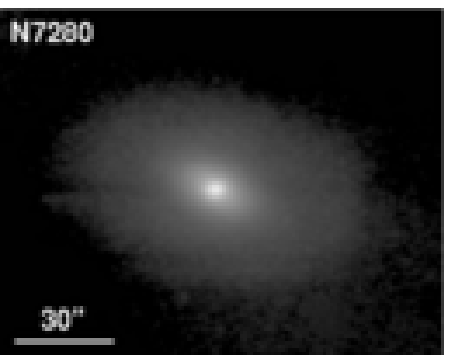}
\includegraphics[keepaspectratio=true,clip=true,width=4cm]{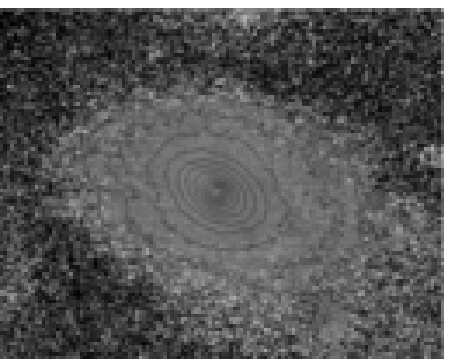}\\
   \includegraphics[keepaspectratio=true,clip=true,width=4cm]{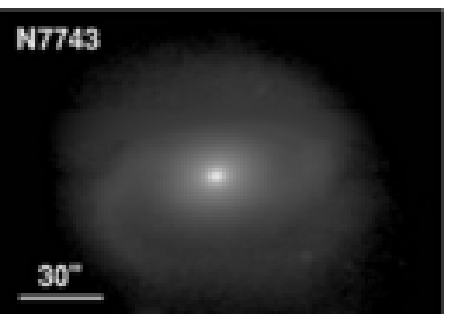}
\includegraphics[keepaspectratio=true,clip=true,width=4cm]{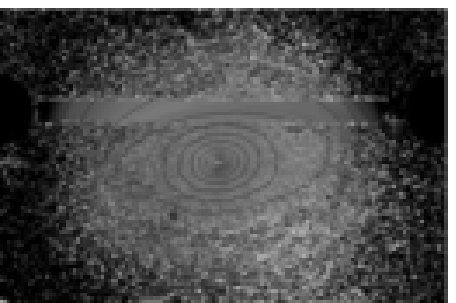}
      \caption{J-band images (left) and J$-$Ks colour maps (right) of the galaxies in our sample.
Colour maps have J-band isophotal contours overlaid and are coded so that redder features are darker.
North is up, East to the left.}
         \label{gals}
   \end{figure*}

\begin{figure*}
   \centering
   \includegraphics[keepaspectratio=true,clip=true,angle=90,width=160mm]{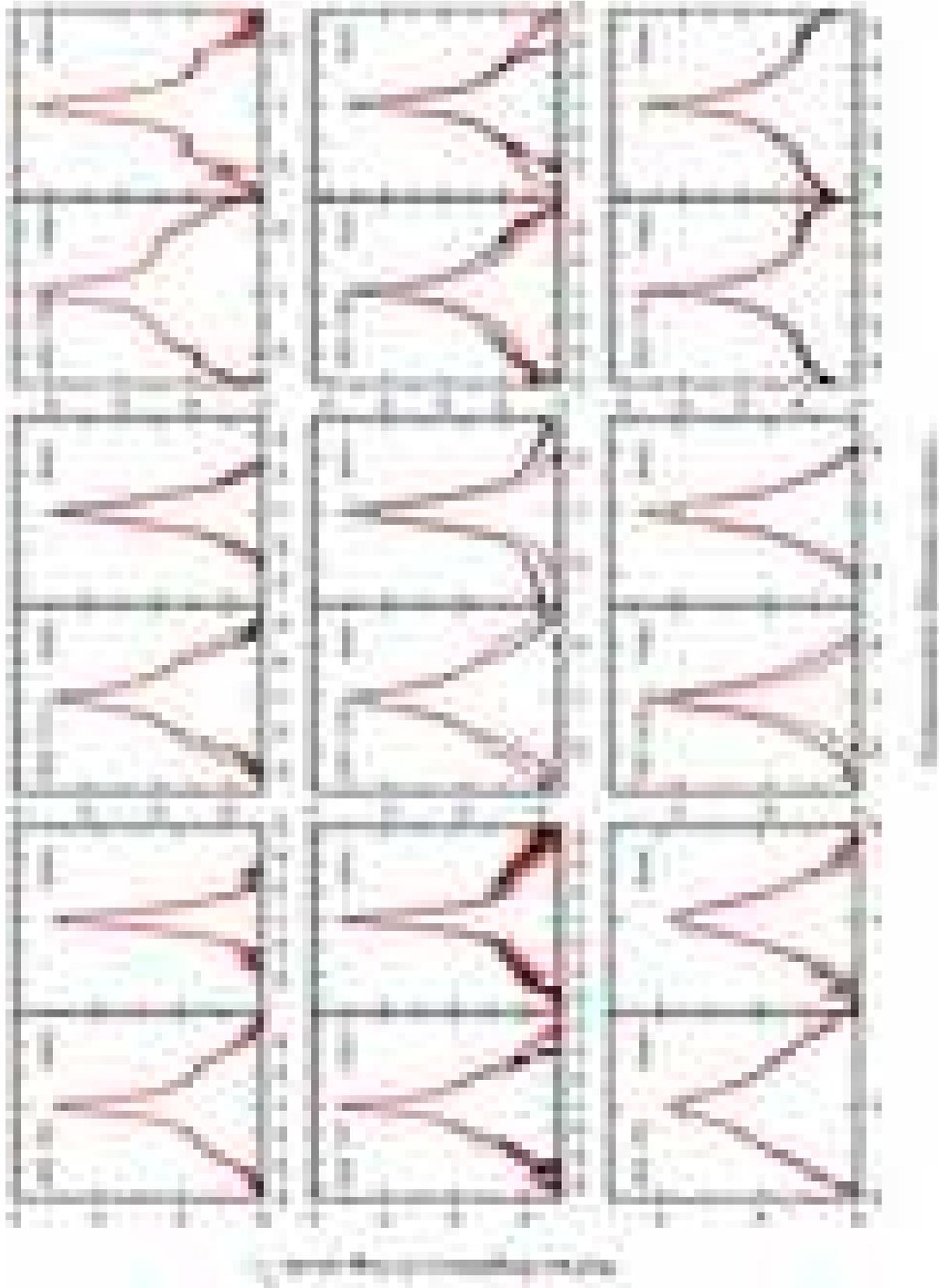}
      \caption{J-band surface brightness profiles along the bar major and minor axes for the
galaxies in our sample. Red lines refer to deprojected images.}
         \label{barprof}
   \end{figure*}

\begin{figure*}
   \centering
   \includegraphics[clip=true,keepaspectratio=true,width=8cm]{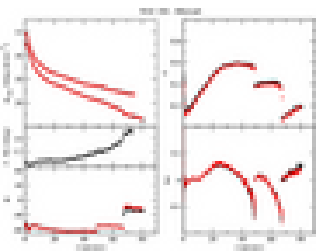}\hspace{0.5cm}
\includegraphics[clip=true,keepaspectratio=true,width=4cm]{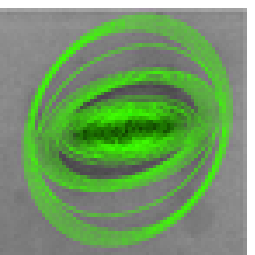}\hspace{0.5cm}
\includegraphics[clip=true,keepaspectratio=true,width=4.05cm]{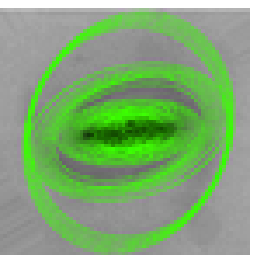}
      \caption{Radial profiles resulting from the ellipse fitting to the isophotes of NGC 266.
Top left: surface brightness in J and Ks. Bottom left: position angle
(from North to East). Top right: ellipticity. Bottom right: the b$_4$ Fourier
coefficient. Middle panel at left: J$-$Ks colour. Red points correspond to the analysis on
deprojected images. J-band images of the galaxy correspond to direct (left) and deprojected (right)
views and have a fraction of the ellipse fits overlaid.}
         \label{res266}
   \end{figure*}

\begin{figure*}
   \centering
   \includegraphics[clip=true,keepaspectratio=true,width=8cm]{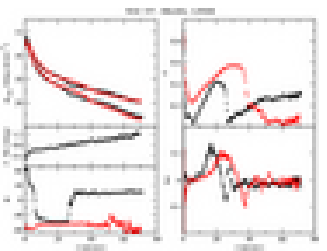}\hspace{0.5cm}
\includegraphics[clip=true,keepaspectratio=true,width=4cm]{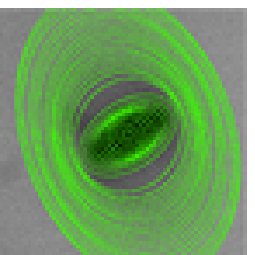}\hspace{0.5cm}
\includegraphics[clip=true,keepaspectratio=true,width=4.05cm]{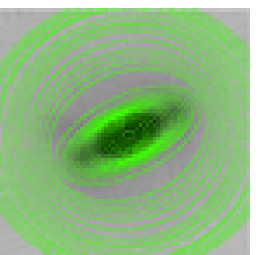}
      \caption{Same as Fig. 3 but for NGC 357.}
         \label{res357}
   \end{figure*}

NGC 7743 is a more ambivalent case. The plateau is of very short
extent and reaches, in the deprojected case, a relatively small value
($\approx0.4$), considerably smaller than that of the other, clear
MH-type bars, whose maximum deprojected ellipticity is, in all cases,
about 0.6. Furthermore, a
closer inspection of the images and ellipse fits of NGC 7743 (Figs. \ref{gals} and
\ref{res7743}) reveals that the peak in the ellipticity profile and the corresponding
steep drop are not caused by the bar, as in the cases discussed above,
but by the inner parts of the
outer spirals arms. All these arguments
taken together lead us to classify NGC 7743 as an MD-type bar,
rather than an MH one. It could also be an intermediate case. In MH
types a considerable amount of angular momentum is exchanged within
the galaxy, emitted by the inner disc and absorbed by the halo and the
outer disc. In MD types less angular momentum is exchanged. NGC 7743
could be an intermediate type, with an intermediate amount of angular
momentum exchanged.

As already mentioned, three other galaxies decidedly deviate from the clear
pattern of MH-type galaxies, namely NGC 799, NGC 1638 and NGC 7280.
Note that these three galaxies have a maximum in the deprojected ellipticity smaller
than that occurring in the other galaxies.
NGC 799 has a plateau of a very short extent, followed by a gradual decrease in
ellipticity, rather than by a steep drop as for the galaxies we discussed
above. As we already noted, this latter property is characteristic of MD-type galaxies
(AM02). Unfortunately, our surface photometry for this galaxy does not
reach very deep (see Sect. 4.3), but a comparison to other images available in NED
reveals that our images miss the faint and blue spiral
arms in the outer disc. Nevertheless, Figs. \ref{gals}, \ref{barprof} and
\ref{res799} argue strongly for an MD-type bar. In particular, in the
region where the position angle of the isophotes stays constant, the ellipticity
first increases and then decreases steadily, a clear signature of an
MD-type bar.
NGC 7280 also has a drop in ellipticity clearly less sharp than those occurring in the MH-type
bars, and our images for this galaxy are amongst our deepest ones. Again, the bar is evident
in the ellipse fits (see Figs. \ref{gals} and \ref{res7280}).

NGC 1638 is a less clear case.
It shows just a big plateau at a relatively low ellipticity value,
with no clear drop. Such a behaviour could be observed in a barred galaxy if the
ellipse fits did not go beyond the end of the bar. Since our fits reach 50
arcsec, and 21.5 and 19.5 mag arcsec$^{-2}$ in the J and Ks images, respectively,
this is rather unlikely. We are more inclined to believe that NGC
1638 is truly {\em not} barred. In fact, the classification from the RC3, SAB(rs)0$^o$?, means there
is a lot of uncertainty, and
no sign of a bar can be seen either on the images or the surface
brightness profiles. The value of the ellipticity is
relatively low and the position angle and b$_4$ profiles also do not
show any clear sign of a bar. We checked optical and near-infrared images available in NED
and also found no sign of a bar. A search in the literature reveals that the
Revised Shapley-Ames catalog \citep{san81} considers this galaxy as unbarred, which
is not very surprising, since these authors were less compelled to classify
a galaxy as barred, in particular as weakly barred, than the classifiers in the
RC3 \citep[see][]{gadphd}. Furthermore, \citet{ebn88} also found no sign of a bar.
We thus conclude that out of the four galaxies which are
not MH-type, one is unbarred, two are clearly of MD-type and the
remaining one is either MD, or intermediate.

\subsection{The b$_4$ Radial Profiles}
\label{sec:b4}

A systematic behaviour in the b$_4$ radial profiles (projected and deprojected)
is also clearly apparent for most of our galaxies.
These profiles generally have small values for the inner roughly 10
arcsec, rise due to the bar, then fall to negative values and finish close to zero. The
values of b$_4$ become negative (indicating boxy isophotes)
roughly at the radius where the ellipticity reaches the plateau or
maximum, or slightly after that. This occurs in the outermost region
of the bar where the influence from the bulge is minimum,
and thus also argues in favour of a rectangular-like shape for bars. In addition, the minimum
in b$_4$ (i.e., the maximum boxyness) in our MH-type galaxies occurs
at a larger distance from the centre than the ellipticity peak,
arguing that this peak gives only a lower limit for the bar
length of MH-type bars. Interestingly enough, the minimum in b$_4$ is located at the same
position as the steep drop in ellipticity (see above), and so this is
another argument in favour of using the location of this drop as a
characteristic scale-length and to compare it to the bar length, at
least for MH-type bars. NGC 266 shows an interesting behaviour that
happens repeatedly in the ellipticity, b$_4$ and position angle profiles. One sees three
distinct regions with the same boundaries in the three profiles. The first region clearly
corresponds to the bar, the second to the oval disc/inner spiral arms just outside the bar (discussed above),
and the third to the outer disc. This can also be assessed through optical images.
To a lesser extent this is also seen in NGC 1211 and NGC 7080. On the other hand,
NGC 799, NGC 7280 and NGC 7743 have the least prominent
values of b$_4$ and, interestingly enough, {\em these are also the bars with the lowest
ellipticity maxima}. This holds for both projected and deprojected measurements and reinforces
our suggestion that NGC 799, NGC 7280 and NGC 7743 are indeed real cases of MD bars.
To fully establish this point, in a future paper we will perform Fourier decomposition of the
galaxies in our sample, since MH and MD bars have also distinct signatures in this kind of analysis,
as mentioned above. The various differences between MH and MD bars are discussed at length in AM02,
including edge-on morphology and kinematics.

\subsection{Colour Profiles}

The colour profiles in Figs. \ref{res266} to \ref{res7743} were built from the ellipse fits in each band separately.
This is justified since the relevant geometrical properties of the fitted ellipses are generally
identical in both bands, meaning that light from different regions of the galaxy is not mixed.
A careful inspection of these profiles reveals varied behaviours.
NGC 799 and NGC 1358 present a fairly flat, or slightly negative, global colour gradient (bluer outwards).
On the other hand, NGC 357, NGC 1638 and NGC 7280 have positive global colour gradients
(redder outwards), although the slope in the latter is small.
NGC 266, 1211, 7080 and NGC 7743 show an inner flat colour profile with a
significant reddening after a certain radius. While the inner parts of these
colour profiles are well estimated, we can not rule out the possibility that a difficult
sky subtraction, in particular in the Ks band,
is the reason behind at least some of the outer red colours. Even
considering all our efforts for a good sky subtraction, this is not a trivial task,
especially when pushing to faint brightness levels in the NIR. We have not found
any relation of this feature with the presence of the moon or non-photometric nights.
To be on a safe side, it is better to consider the observed sudden outward reddening as spurious.

With the exception of NGC 7080, all galaxies have nuclei that are bluer
than their immediate surroundings, and many times the nucleus is the bluest part of the galaxy
(see also the colour maps in Fig. \ref{gals}).
This does not seem to be strictly related to AGN activity, although this connection is
difficult to analyse since we do not have information on AGN activity for all our galaxies (see
Table 1). As pointed out by \citet{gad01} these colour variations probably reflect changes
in the {\em age} of the stellar population. In particular, the blue nuclei observed here are
likely to be the result of starbursts fuelled by secular processes induced by the bars.
Apart from the outer and innermost regions the colour profiles are generally quite
flat.

\section{Characteristic Scale-Lengths Related to the Bar Size}

\subsection{Bar Lengths from Projected and Deprojected Images}

Measuring the bar length is not a trivial problem, as thoroughly
discussed by \citet{erw05} and by AM02 for bars in real galaxies and in
$N$-body simulations, respectively. Further discussion of the ways to
measure the length of $N$-body bars have been given by
\citet{O'neill}, \citet{mic06} and \citet{MVSH}. AM02 introduced seven
different characteristic scale-lengths and discussed their use for
measuring the bar lengths in MH-type and MD-type bars.
Here we will broadly follow their lead, since our ultimate
goal is a comparison of observed and $N$-body bars. We
will thus adopt four of their seven measures and add three more. It should
be noted that there are small differences between the ways the
measurements are made in the observations and in the simulations, but
they are deemed negligible for the purposes of our comparisons.

\begin{figure*}
   \centering
   \includegraphics[clip=true,keepaspectratio=true,width=8cm]{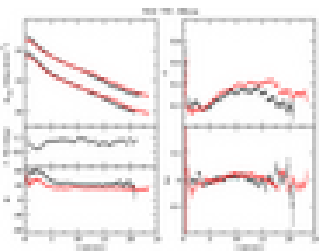}\hspace{0.5cm}
\includegraphics[clip=true,keepaspectratio=true,width=4.05cm]{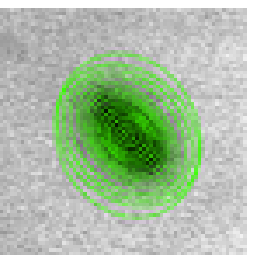}\hspace{0.5cm}
\includegraphics[clip=true,keepaspectratio=true,width=4cm]{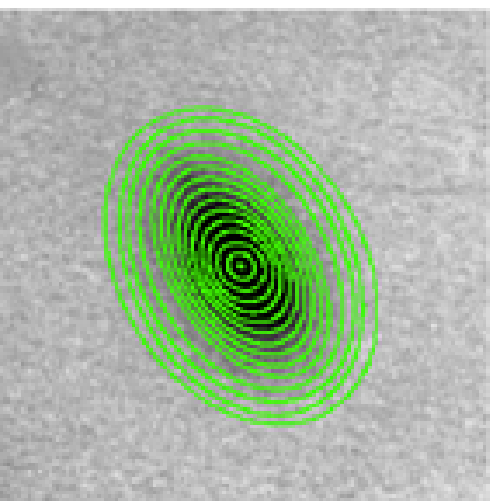}
      \caption{Same as Fig. 3 but for NGC 799.}
         \label{res799}
   \end{figure*}

\begin{figure*}
   \centering
   \includegraphics[clip=true,keepaspectratio=true,width=8cm]{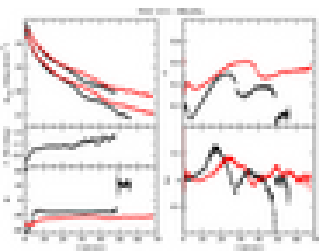}\hspace{0.5cm}
\includegraphics[clip=true,keepaspectratio=true,width=4cm]{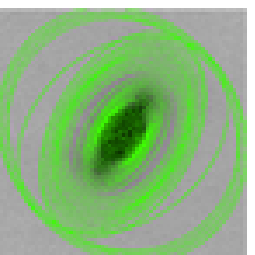}\hspace{0.5cm}
\includegraphics[clip=true,keepaspectratio=true,width=4.15cm]{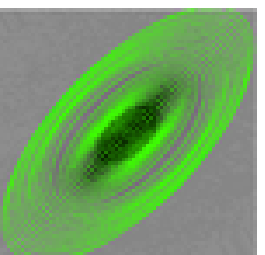}
      \caption{Same as Fig. 3 but for NGC 1211.}
         \label{res1211}
   \end{figure*}

\begin{figure*}
   \centering
   \includegraphics[clip=true,keepaspectratio=true,width=8cm]{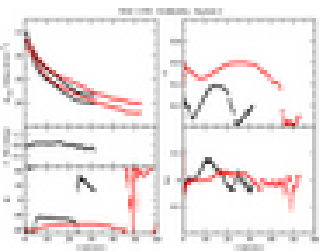}\hspace{0.5cm}
\includegraphics[clip=true,keepaspectratio=true,width=4cm]{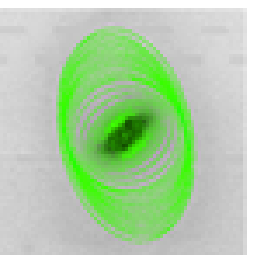}\hspace{0.5cm}
\includegraphics[clip=true,keepaspectratio=true,width=4.1cm]{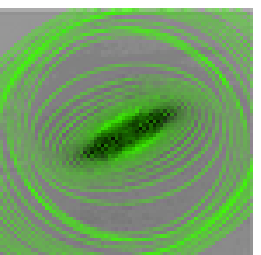}
      \caption{Same as Fig. 3 but for NGC 1358.}
         \label{res1358}
   \end{figure*}

\begin{figure*}
   \centering
   \includegraphics[clip=true,keepaspectratio=true,width=8cm]{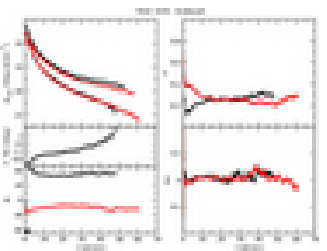}\hspace{0.5cm}
\includegraphics[clip=true,keepaspectratio=true,width=4.35cm]{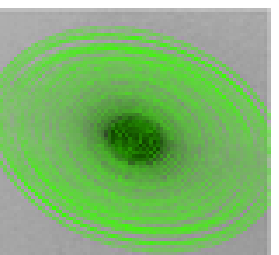}\hspace{0.5cm}
\includegraphics[clip=true,keepaspectratio=true,width=4cm]{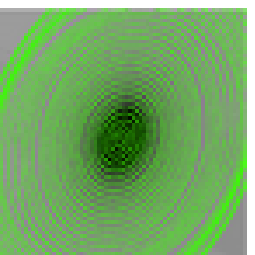}
      \caption{Same as Fig. 3 but for NGC 1638.}
         \label{res1638}
   \end{figure*}

\begin{figure*}
   \centering
   \includegraphics[clip=true,keepaspectratio=true,width=8cm]{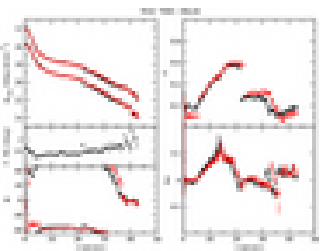}\hspace{0.5cm}
\includegraphics[clip=true,keepaspectratio=true,width=4cm]{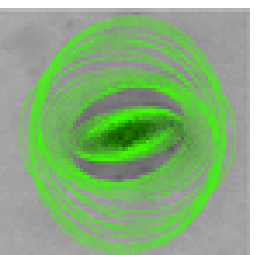}\hspace{0.5cm}
\includegraphics[clip=true,keepaspectratio=true,width=4.05cm]{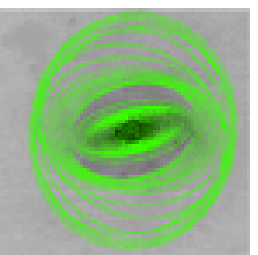}
      \caption{Same as Fig. 3 but for NGC 7080.}
         \label{res7080}
   \end{figure*}

\begin{figure*}
   \centering
   \includegraphics[clip=true,keepaspectratio=true,width=8cm]{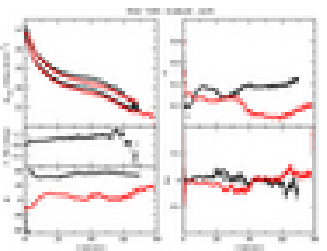}\hspace{0.5cm}
\includegraphics[clip=true,keepaspectratio=true,width=4cm]{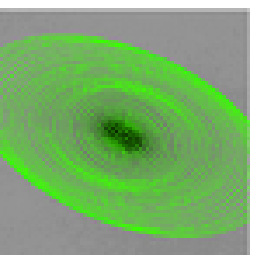}\hspace{0.5cm}
\includegraphics[clip=true,keepaspectratio=true,width=4.05cm]{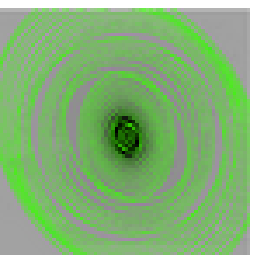}
      \caption{Same as Fig. 3 but for NGC 7280.}
         \label{res7280}
   \end{figure*}

\begin{figure*}
   \centering
   \includegraphics[clip=true,keepaspectratio=true,width=8cm]{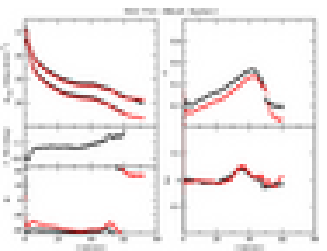}\hspace{0.5cm}
\includegraphics[clip=true,keepaspectratio=true,width=4.15cm]{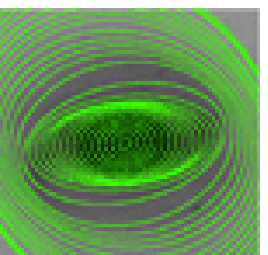}\hspace{0.5cm}
\includegraphics[clip=true,keepaspectratio=true,width=4cm]{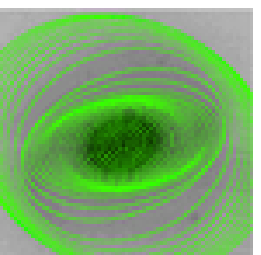}
      \caption{Same as Fig. 3 but for NGC 7743.}
         \label{res7743}
   \end{figure*}

The phase angle of the $m=2$ Fourier component should be constant
in the bar region and so we can measure
the bar length, $L_{{\rm phase}}$, from the position where the phase angle changes
abruptly at the transition between the bar and the outer disc
(or another component such as a ring, or spiral arms).
In practice, we will determine the radius at which the
position angle of the fitted ellipses changes by more than
$10^{\circ}$ from that of the bar. The latter is defined as the average
position angle of the ellipse fits to the isophotes in the bar region, i.e.,
the region within which no significant change ($>10^\circ$) in the position angle radial
profile occurs (see Figs. \ref{res266} to \ref{res7743}). Ten degrees is a usually chosen, yet
arbitrary, threshold, that results in fair estimates in most cases.

From the ellipticity profiles we can obtain three characteristic
scale-lengths linked to the bar.
Namely, the position of the maximum ellipticity,
$L_{\epsilon_{\rm max}}$ ($L_{b/a}$ in AM02),
the position of the steep drop in ellipticity, $L_{{\rm drop}}$ (more
precisely, the last position before the maximum change in slope in
the ellipticity profile just after $L_{\epsilon_{\rm max}}$), and the
position of the first ellipticity minimum outside the bar's maximum
ellipticity, $L_{\epsilon_{\rm min}}$. This last characteristic length
was not included in those used in AM02, but has been later introduced
and found useful \citep{erw03,erw05}. As already discussed in AM02,
for galaxies of the MH-type, whose ellipticity profile has a plateau, the
measure $L_{\epsilon_{\rm max}}$ is neither meaningful nor useful,
since there is no significant difference in ellipticity
to distinguish one point of the plateau from another. Similarly,
$L_{{\rm drop}}$ is not meaningful for galaxies of the MD-type, since their
ellipticity profile has no steep drop. Finally, $L_{\epsilon_{\rm min}}$ should be
roughly the same as $L_{{\rm drop}}$ in the MH-type galaxies, since they have a steep drop.

A further measure of the bar length can be obtained from the
surface brightness profiles. Usually, the end of the bar is taken as
the end of the flat ledge, due to the bar, along the bar major axis
profile (see Fig. \ref{barprof}). We will call this scale-length $L_{{\rm prof}_i}$.
But one can also define the end of the bar as the position where
the drop after the flat ledge in the brightness profile joins up to
the disc. We will show below that the latter definition generally gives better
estimates, and we will call it $L_{{\rm prof}_f}$. As already
discussed in AM02, these measurements can not be applied to MD-type
galaxies, since these do not have a ledge in the photometric profiles.
Furthermore, sometimes these measurements are difficult to obtain,
even for MH-type bars, if the ledge is not clearly defined. In particular,
$L_{{\rm prof}_f}$ is usually more difficult to obtain than $L_{{\rm prof}_i}$,
demanding a higher S/N. In some cases, it could only be clearly defined in the elliptically
averaged profiles in Figs. \ref{res266} to \ref{res7743}.
Nevertheless, we find that, for the cases where these measurements can be applied,
the mean difference between the measurements at the two opposite sides of the bar is
only about 1--2\%.

To these scale-lengths we will add yet another one, obtained from the
b$_4$ radial profiles, namely the position of the b$_4$ peak,
$L_{{\rm b_4}}$. Since b$_4$ should be positive in the regions where
the bulge contribution is important and negative outside it, we expect
$L_{{\rm b_4}}$ to be smaller than the bar length.

Note that any of the scale-lengths mentioned might fail to reveal the true
length of the bar in certain cases. For instance, if the bar is smoothly connected with
another component, such as a lens or spiral arms, with a similar position angle or
ellipticity, parameters like $L_{{\rm phase}}$ or $L_{\epsilon_{\rm max}}$
can be significantly larger than the length of the bar. Inspecting the images is thus a
mandatory safety check, especially for galaxies with a complex morphology.

We measured these characteristic scale-lengths for all our galaxies
and give the results for the projected and deprojected images in Tables
3 and 4, respectively. We give the results both in arcsec, so as to
allow comparisons with Figs. \ref{gals} to \ref{res7743}, and in Kpc, to allow comparisons
between different galaxies and with previous work. In Sects.
\ref{sec:ellipticity} and \ref{sec:b4} we discussed the existence of a
second separate sub-structure in the ellipticity and b$_{4}$ radial
profiles of NGC 266, NGC 1211 and NGC 7080, presumably due to a lens-like or oval
disc component. We measured the characteristic scale-lengths of these
sub-structures too and include the results in Tables 3 and 4, under the
entries NGC 266b, NGC 1211b and NGC 7080b, respectively.

Note that, as expected, some of the measurements are just not doable. For instance,
one can not rigorously define $L_{{\rm drop}}$ in the MD-type bar of NGC 799.
Furthermore, $L_{\epsilon_{\rm max}}$ in MH-type galaxies like NGC 266 does not represent
true peaks in ellipticity, but just the position in the ellipticity plateau that
happens to have a slightly larger value. Some measurements might be complicated in the
deprojected images. While in the direct
images of NGC 799 and NGC 1211 $L_{{\rm phase}}$ is clearly defined, it is
not in the corresponding deprojected images. In NGC 799, the change in the position angle
of the ellipse fits in the direct image is small and might have been just smeared
out in the deprojection of the image. In NGC 1211, since the inclination angle is relatively large,
$L_{{\rm phase}}$ is pushed to the limits of the image, as mentioned in Sect. \ref{sec:ellipticity}.
For both galaxies, we assume that the deprojected radial profiles just reach $L_{{\rm phase}}$ and
the corresponding values are put within parenthesis in Table 4. Comparing the profiles of the
projected and deprojected images (Figs. \ref{res799} and \ref{res1211}) shows that this is a reasonable
assumption. For instance, the ratio of $L_{\epsilon_{\rm max}}$ in the deprojected and direct images
is of the same order as the corresponding ratio for $L_{{\rm phase}}$ if one uses our assumption.

The position angle threshold in the definition of $L_{{\rm phase}}$ is another issue. If the
position angles of the bar and the component just outside it are roughly the same, then
$L_{{\rm phase}}$ might not be very useful. Indeed, $L_{{\rm phase}}$ does not mark
the transition between the bar and the next component if their position angles are not
more than 10$^\circ$ apart. This is the case in the direct and
deprojected images of NGC 1211, and in the direct image of NGC 7280. In fact, Table 3
shows that $L_{{\rm phase}}$ is considerably bigger than $L_{\epsilon_{\rm min}}$
only in these galaxies. In NGC 1211, the oval disc has a position angle very similar to that of
the bar, and in NGC 7280 the outer disc has a position angle similar to that of the bar. Hence,
in these cases, the arbitrary 10$^\circ$ threshold that defines $L_{{\rm phase}}$ does not
give a meaningful result, and one needs to search for a less conspicuous change
in position angle that might indicate the transition between the components.
A close inspection of the position angle radial profiles of NGC 1211 (Fig. \ref{res1211}) reveals
glitches at 27 arcsec and 37 arcsec in the direct and deprojected measurements, respectively,
and we will hereafter use these values as $L_{{\rm phase}}$.
In the position angle radial profile of the direct image of NGC 7280 it is easy to notice
the transition between the bar and the outer disc starting at 21 arcsec from the centre, but it takes
11 arcsec more for this change to reach the threshold of 10$^\circ$ (see Fig. \ref{res7280}). Hence, in the
following, we will use $L_{{\rm phase}}=21$ arcsec for the direct
image of NGC 7280. It is straightforward to verify that these values are a much
better estimate of the bar length in these galaxies than the values taken directly from the nominal
definition of $L_{{\rm phase}}$.

The values of the bar length in NGC 7743 quoted in Table 3 seem all to be excessively large
in a comparison with the galaxy image in Figs. \ref{gals} and \ref{res7743}. In particular,
as mentioned in the last section, $L_{\epsilon_{\rm max}}$ seems to happen outside the bar, as a result of
the smooth joining of the isophotes of the bar and the spiral arms. \citet{erw05} argues
similarly. The other radial profiles also show smooth transitions.
However, both the projected and deprojected ellipticity profiles have a small glitch at 33 and
35 arcsec from the centre, respectively. Comparison with the images argues that
this glitch might be caused by the transition between the bar and the spiral arms, and so
we choose its position as our fiducial value of the bar length in this galaxy. This is in agreement
with the estimates of \citet{erw05} that go from 31 to 37 arcsec. Nevertheless, since no
glitches are present in the position angle profiles, we have to discard NGC 7743 in some
of the analysis below.

Apart from NGC 7743, there are no published estimates of the bar lengths of the galaxies in our sample.
However, the values we obtain are of the same order as generally found in studies using optical
images \citep[see, e.g.][and references therein]{erw05,gad06}.

\begin{figure*}
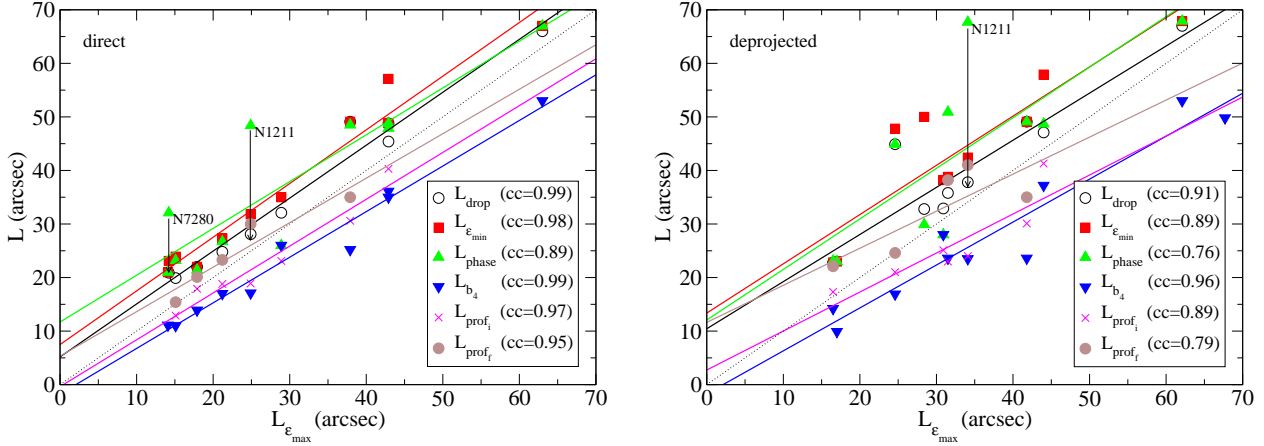

   \centering
   \includegraphics[keepaspectratio=true,clip=true,width=8cm]{scl.eps} \hskip 0.5cm
\includegraphics[keepaspectratio=true,clip=true,width=8cm]{sclfo.eps}
      \caption{Correlation between all measured scale-lengths in direct (left) and deprojected
images (right), using $L_{\epsilon_{\rm max}}$ as reference.
Solid lines are linear fits to the data, for each scale-length separately, colour-coded
as indicated. The dotted line is a one-to-one correspondence line. The arrows indicate our new definitions
of $L_{{\rm phase}}$ for NGC 1211 and NGC 7280. Correlation coefficients are also shown. As expected,
all scale-lengths are clearly correlated. Note also that the uncertainties in image deprojection rise the spread
in the correlations.}
         \label{scls}
   \end{figure*}

The two panels in Fig. \ref{scls} show that all measured scale-lengths are strongly correlated.
This is not surprising, since they refer to a specific structural component, which is,
for most of the points in these plots, the bar. These plots also include the oval discs measurements
in NGC 266, NGC 1211 and NGC 7080, and the scale-lengths of NGC 7743.
Interestingly, these structures follow the
same relation as the bars. This figure also shows that, as expected, $L_{{\rm b_4}} < L_{\epsilon_{\rm max}}$.
In Tables 3 and 4 one sees in fact that $L_{{\rm b_4}}$ is almost always the smallest value.
In addition, it is interesting to note that $L_{{\rm prof}_i}$ is usually similar to $L_{{\rm b_4}}$.
Note also that the spread in the correlations is larger in the deprojected measurements, most likely
due to the uncertainties in image deprojection.

In order to compare the results of the various methods to determine the bar length we need to use
one of the characteristic lengths as a yard-stick. As discussed above,
$L_{{\rm b_4}}$ is considerably smaller than the bar
length, while $L_{{\rm drop}}$, $L_{{\rm prof}_i}$ and $L_{{\rm prof}_f}$ are not meaningful for
MD bars, and $L_{\epsilon_{\rm max}}$ is not meaningful for MH bars. This leaves
$L_{{\rm phase}}$ and $L_{\epsilon_{\rm min}}$. We arbitrarily choose the
former and give the results of the bar length relative to this measure
in Tables 5 and 6 for observed and deprojected images, respectively.
Note that, since these are scale-length ratios, defined on the same position
angle, one can work directly with the values from the
projected images, assuming that these ratios do not change in the
deprojected images. In fact, since the values in Tables 5 and 6 depict similar trends, we can
carry on with our analysis using only the data in Table 5.
The fact that there are a few differences means that image deprojection
might sometimes modify the various scale-lengths by different factors.
To assess whether these differences are spurious, or not, one has to
consider the morphology of each component in the galaxy
and its true position in space with respect to the plane of the sky.
This is beyond our present scope. Note that, for NGC 1211 and NGC 7280, the values of
$L_{{\rm phase}}$ used in Tables 5 and 6 are not the nominal ones, as in Tables 3 and 4,
but those defined and discussed above after a closer inspection of the position
angle radial profiles.

Let us now compare the values of the various scale-lengths as given by
Table 5, i.e., after scaling with $L_{{\rm phase}}$.
The smallest one is clearly $L_{{\rm b_4}}$, with a mean
value of $0.58\pm0.07$. By definition, this quantity is expected to
be smaller than the bar length, since the maximum of b$_4$ should
occur in the part dominated by the bulge, not at the end of the
bar. $L_{{\rm prof}_i}$ is the second smallest scale-length, with a mean
value of $0.68\pm0.1$. As already discussed,
$L_{\epsilon_{\rm max}}$ is meaningful for MD-type bars, for which we find
a mean value of 0.68; there are, however, only two such galaxies for which we
could do this analysis. If
we include the MH-types in the statistics we find $0.76\pm0.1$, but in
many cases this value is ill defined. $L_{{\rm prof}_f}$ comes next, with
a mean value of $0.86\pm0.18$. As expected, because this is a difficult
measurement, the standard deviation is larger than for other scale-length
measurements. Nonetheless, it gives
an estimate closer to $L_{{\rm phase}}$ than the one given by $L_{{\rm prof}_i}$
and is in fact consistent with it to within the errors.
We are left with two scale-lengths,
$L_{\epsilon_{\rm min}}$ and $L_{{\rm drop}}$. The latter is well defined
only for MH-types, for which we get a value of $1.00\pm0.04$. This
value hardly changes if we add the one MD-type galaxy for which this
measurement is possible. For $L_{\epsilon_{\rm min}}$ we get a mean
value of 1.05 with a standard deviation of 0.06. Our measurements thus
show that four determinations -- namely $L_{\rm phase}$,
$L_{\epsilon_{\rm min}}$, $L_{{\rm drop}}$ and to a lesser extent $L_{{\rm prof}_f}$
-- give values which are equal within the
errors. Although our result needs to be confirmed both with
a larger observational sample and with a sample of $N$-body bars, it
is tempting to conclude that these four determinations, or their
average, should allow a fair measurement of the bar length.

Nonetheless, as we already noted above, $L_{{\rm drop}}$ coincides with the position
where the minimum in b$_4$ happens. In fact, in all cases where this could be
defined, i.e. the MH-type bars, both radii are identical. This holds for projected and
deprojected measurements. The only exception occurs in the deprojected image of NGC 1358,
but below we will show evidences that suggest that the deprojected measurements for this galaxy
might be to some extent affected by uncertainties in its values of $i$ and PA$_{\rm ln}$.
Since the position of the minimum in b$_4$
is the position where the bar reaches its maximum boxyness, this argues strongly
in favour of $L_{{\rm drop}}$ to define the end of the bar, in particular because b$_4$
reaches zero very quickly after that.

\subsection{Analytically Deprojected Bar Lengths}

Because of image stretching, the use of deprojected images to obtain deprojected bar lengths is
subject to spurious geometric effects and errors, in particular if the inclination angle is large.
However, theoretical studies on the orbital structure of barred galaxies and
$N$-body simulations predict that the inner parts of bars might be vertically thick but the outer
parts of bars are vertically thin \citep[see][]{ath05b}. This structure
was confirmed for our own galaxy \citep[and references therein]{MWbar},
for M31 \citep[][see also \citealt{ath06b}]{ath06a} and for a sample of
30 edge-on disc galaxies observed in the NIR \citep{BAADBF06}.
If this is generally confirmed, then by
knowing the inclination angle of the galaxy, and the position angles of the bar and of the line of
nodes, one can obtain an analytical expression for the true bar length from its projected size.
This would work in the majority of cases; the
only exceptions being near edge-on galaxies, or highly inclined
galaxies with the bar major axis near the galaxy minor axis. In such
cases and if the vertical extent of the inner part of the bar is sufficient, it
could block the end of the bar from our view, so that both image and analytical
deprojections would overestimate the bar length.

We can first assume that a bar can be treated as a line and call this assumption our 1D approximation.
Thus, with simple trigonometry arguments, it is straightforward to show that the true, deprojected length
of the bar is given by

\begin{equation}
L_{\rm bar}=L_p\left(\sin^2\alpha \, \sec^2i + \cos^2\alpha\right)^{1/2},
\end{equation}

\noindent where $L_p$ is the observed, projected bar length, measured as, e.g.
$L_{{\rm drop}}$, $\alpha$ is the difference between the position angle of the line of
nodes and the position angle of the bar, and $i$ is the inclination angle
\citep[see also][]{mar95}. As expected,
a bar parallel to the line of nodes shows always its true length, regardless of the
inclination angle. Of course, this equation diverges when
$i$ reaches 90$^\circ$, i.e., in the case of perfectly edge-on galaxies.

But bars are not thin lines and so it is unclear whether our 1D approximation holds for
real bars, especially those which are not very narrow.
Thus, in the Appendix, we derive expressions for the deprojected semi-major and semi-minor axes, as
well as position angle, of an ellipse seen in projection. This allows us to obtain analytically
measurements of the lengths, ellipticities and position angles of the deprojected bars in our
galaxies, taking into account the 2D properties of the bars\footnote{The source code of a
{\sc fortran} program to perform these calculations can be downloaded
at http://www.mpa-garching.mpg.de/$\sim$dimitri/deprojell.f .}.

We calculated $L_{\rm bar}$ using Eq. (1) and the expressions from our 2D treatment
and show the results in Table 7. The adopted values
of the inclination angle, position angle of the line of nodes and
$\alpha$ are taken from Table 1, whereas the values for the position
angle of the bar and $L_p$ are those extracted from our images. We used
$L_p=L_{{\rm drop}}$ for our MH-type galaxies and $L_p=L_{{\rm phase}}$ for our MD-type
galaxies, for which $L_{{\rm drop}}$ is not defined. The 1D and 2D deprojected bar lengths
agree very well with each other if the inclination is not too high. However, since the 1D approximation
does not consider changes in the position angle, the values it
provides for the bar length are
always slightly smaller. Table 7 also shows the deprojected
ellipticities and position angles obtained with the 2D analysis. Comparing them with what
is obtained from the deprojected images in Figs. \ref{res266} to \ref{res7743} one sees very good agreement.
This means that the expressions in the Appendix can be reliably used to obtain deprojected ellipticities
and position angles.

\begin{figure*}
   \centering
   \includegraphics[keepaspectratio=true,clip=true,width=8cm]{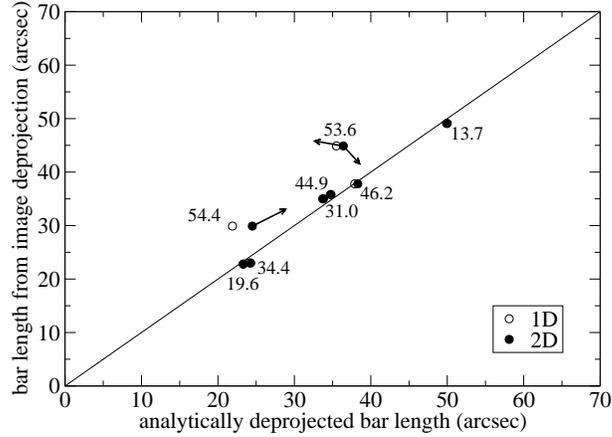}
      \caption{The true bar length as measured from the deprojected images plotted against
the same quantity as estimated analytically. Empty circles refer to our 1D approximation while filled circles
correspond to the equations derived in the Appendix in the 2D analytical treatment.
The inclination angle $i$ of each galaxy from Table 1 is written
next to the corresponding points, and the solid line depicts a one-to-one correspondence.
The arrows indicate values for NGC 1358 and NGC 7280 assuming an uncertainty
of $\pm20^\circ$ in PA$_{\rm ln}$. In these cases, which are the most inclined galaxies in our
sample, the discrepancy between the results from the 2D analytical treatment and the deprojected
images is alleviated if the true PA$_{\rm ln}=35^\circ$ for the former and PA$_{\rm ln}=98^\circ$
for the latter. Note that assuming for the latter a value for PA$_{\rm ln}=58^\circ$
produces only a very weak oval whose length is
not clearly discernible in the radial profiles from ellipse fits.
This plot shows clearly that reliable estimates for the bar length can be obtained
analytically.}
         \label{lbanl}
   \end{figure*}

In Figure \ref{lbanl} we show that $L_{\rm bar}$ from both the 1D and 2D analyses agree very well with
each other and also with estimates from deprojected images, but only when the inclination angle is smaller
than about 50$^\circ$. Two of our galaxies do not satisfy this criterion: NGC 1358 and NGC 7280;
the length of their bars, as estimated from deprojected images, is overestimated by 8.5 arcsec and
5.4 arcsec, respectively, i.e. $\approx20\%$ in both cases. Let us examine
closer these two exceptional cases. How sure are we
that the inclination angles we applied are correct? For NGC 1358, this parameter, as quoted
in the literature, ranges from 37$^\circ$ to 55$^\circ$ (see results in NED and LEDA).
Our choice is closer to the latter value and the ellipticity profile of the deprojected
image argues favourably to it, as the ellipticity is close to zero in the outer parts (Fig. \ref{res1358}).
To be reassured of that, we created a deprojected image of NGC 1358 assuming that
$i=40^\circ$ and found that the disc remains significantly eccentric, again arguing that the true
inclination angle of this galaxy is closer to 50$^\circ$ than to 40$^\circ$.
Thus, although a smaller inclination angle can alleviate the
discrepancy in the bar length estimates, we believe that our adopted
values are more correct.
For NGC 7280, the values of $i$ in the literature range from 44$^\circ$ to 59$^\circ$.
We chose a value closer to the latter and again a similar analysis as done with NGC 1358
favours our choice. Nonetheless, the wide range of estimates seen in the literature is
an indication of how difficult and uncertain the measurement of $i$ can be.

Alternatively, this discrepancy could be the result of an uncertainty of
$\pm20^\circ$ in PA$_{\rm ln}$. As depicted by the arrows in
Fig. \ref{lbanl}, if the real position angle was 35$^\circ$ the
difference in the case of NGC 1358 could be considerably
alleviated. Similarly, the difference is less pronounced
in the case of NGC 7280 if we add $20^\circ$ to PA$_{\rm ln}$.
Note, however, that assuming for the latter a value for PA$_{\rm ln}=58^\circ$
(i.e. 20$^\circ$ less than our original estimate) produces in the
deprojected image only a very weak oval, whose length is
not clearly discernible in the radial profiles from ellipse fits. To
summarise, we do not believe that the discrepancies are due to
erroneous choices of the viewing angles, but to the difficulty of
obtaining a correct deprojected image in cases that the inclination
angle is considerable. Nevertheless, a larger sample is necessary to
study this issue more thoroughly.

These results show that the deprojected length of bars can
be reliably determined analytically, with no need of using deprojected images.
Furthermore, the fact that the analytical
expressions hold so well means that the assumption that the outer parts of
bars are vertically thin (as opposed to their inner parts), as
predicted by orbital structure work and by simulations,
is correct. Indeed, if the ends of the bar were thick, the image and analytical
deprojections would be significantly discrepant, and Fig. \ref{lbanl}
shows that this is not the case, not even, in a clear way, for our two most inclined
galaxies.

Using the results from our 2D analysis, one sees that, as expected
from the results of the $N$-body simulations in AM02,
MH-like galaxies have longer bars on average than MD-like galaxies:
the median values are, respectively, 8.3 Kpc and 4.0 Kpc. Nevertheless,
to establish this difference fully, with a better statistical weight,
we would need a bigger sample, particularly for MD-type bars.

\section{Discussion}

\subsection{The Effects of Deprojecting Images}

A deprojected image of an inclined galaxy is necessarily approximate,
unless the inclination angle is negligible and/or the galaxy is razor
thin. Indeed, one has to take into account the complete geometry
of the different galaxy components, including the vertical properties
of bulges, discs and bars, and these are not known exactly.
For instance, since we do not know exactly the geometry of the bulge, in particular
its vertical mass distribution and its orientation
with respect to the disc, the stretching performed to generate the deprojected image might produce
spurious effects in the bulge region, especially in the case of galaxies with massive bulges.
One possible solution to this problem, proposed by e.g. \citet{lau04}, is to obtain a model for
the bulge, remove it from the image, and put it back {\em after} deprojection. Yet this approach is
not flawless either since it
assumes that the bulge mass distribution is spherically
symmetric. We prefer to give the results from deprojected images as
indicative only. In a future paper, when we compare our results
to $N$-body bars we will project the $N$-body snapshots rather than use results
from deprojected images.

Nevertheless, deprojected measurements have been used in the literature
and thus it is interesting to examine the changes
introduced in the galaxy images by deprojection \citep[see also][]{jun97}.
One sees in Figs. \ref{res266} to \ref{res7743} that,
when $i$ is smaller than about $30^\circ$, projected and deprojected images are very similar.
Thus, let us now focus our analysis on galaxies with $i$ larger than about $30^\circ$.
One sees that the position of the plateau/peak in ellipticity in their deprojected images moves outwards
in all cases, indicating longer bars. This is expected since, unless the bar is aligned parallel
to the line of nodes, the bar length will always be shortened by projection.
In most cases, the maximum in ellipticity occurs at higher ellipticity values, meaning more eccentric
bars. This of course is due to the fact that the position angle of the bar is
nearer to perpendicular to the galaxy major axis than along it. On the
contrary, in NGC 7280 the peak in ellipticity occurs
at lower values when the image is deprojected, because
the position angle of the bar is closer to the galaxy major
axis. Similar arguments can be applied to $L_{{\rm b_4}}$, clearly defined
for MH-type galaxies, which also moves outwards, while the value of the
b$_4$ peak itself is lowered when projection effects are considered.

The effects of deprojection on the brightness profiles are also clear.
As expected, the isophotes reach larger distances. In addition, the
breaks where the bar ends consistently
move outwards and might also look milder. This is a result of
the amplification of the image in the direction perpendicular to the line of nodes.
Furthermore, there are changes in the surface brightness levels proper, in particular in the centre,
to account for the fact that the area on the sky comprised by the galaxy is larger in the
deprojected image. This highlights the importance of performing a flux-conserving stretching of
the image during deprojection.

Deprojecting the images has also interesting effects
on the position angle radial profiles. In NGC 357 one sees that the outer change in position
angle due to the ring practically disappears. The line of nodes lies roughly along
the position angle of the ring, which is then stretched along the
perpendicular direction. The result is that from being almost perpendicular to the bar,
the ring becomes almost parallel. This is interesting since both
simulations \citep{sch79,sch81}
and analysis of observations \citep{but86,but95} show that inner rings
are preferentially aligned with bars. So at least part of those which are not seen parallel might
be in fact a result from projection effects. Since the ring in the deprojected image is close to
circular, the position angle profile is very noisy in this region. Note also that the difference
between the position angle of the bar and the line of nodes is always larger when the galaxy is
deprojected.

\subsection{Comparing Real and $N$-Body Bars}

Obviously, a successful modelling of the origin and evolution of bars in galaxies
has to provide bars with properties that match those of real barred galaxies.
Conversely, the observation of bars with different properties, when linked
to theoretical studies, may provide clues to explain the observed diversity.
The analysis we present in this paper suggests a number of useful comparisons
to $N$-body realizations of barred galaxies.
An example of how useful such comparisons may be can be seen in \citet{gad03},
where, for instance, the ellipticity profile is used to evaluate how different
$N$-body models compare to a real galaxy.

From the behaviour of the
radial profiles of ellipticity presented here one is able to distinguish cases which are similar
to one of the proto-typical models in AM02. The weak bars in NGC 799 and NGC 7280 share
a similar ellipticity profile ({\em with a gentle drop after the bar}) with their MD models, which indeed
produce weak bars as a result of the limited bar-halo interaction. On the other hand,
the {\em steep drop} in ellipticity seen in our strong bars (in particular,
NGC 266, NGC 1211 and NGC 7080) is a property of the MH models,
that form the strongest bars from the vigorous bar-halo interaction.
Therefore, the abruptness of this drop seems to be a useful indicator to separate real instances
of the MH and MD cases.

One also expects the models to explain the observed lengths of bars.
The lengths of bars, in connection to their ellipticities and Fourier even components,
are related to their strength and importance in the overall evolution of the galaxy.
Other studies suggest in addition that bars can get longer during
the course of their evolution \citep{ath03}, which is also in agreement with recent
observational results \citep{gad05,gad06}. This adds relevance, but also complexity,
to a comparison between the lengths of observed and simulated
bars. Here we aimed at the first step of this comparison, namely how
to measure the bar length.

We postpone to a future paper a more thorough comparison between the observed properties
presented here and those obtained in a similar way from snapshots of $N$-body realizations.

\section{Summary and Concluding Remarks}

In this paper, we made a thorough analysis of morphological and
photometrical properties of a sample of barred galaxies, to be
compared in future work to the corresponding properties of bars in
$N$-body simulations. Our sample is relatively
small, nine galaxies, but this has proven to be an asset since it
allowed us to examine each case separately, in depth, without having to
rely on an automatic treatment. This became particularly clear when we
worked on the scale-lengths. The difficulty in measuring the bar
length highlights the necessity of inspecting each case individually
and making judgements which no automated approach could make. This was
possible here only due to the relatively restricted size of our sample.

We have used two NIR wavelengths, so
that we can follow the properties of the old stellar population, which
contributes most of the visible matter. Surface brightness radial profiles,
obtained either from cuts along the bar major and minor axes, as well as
globally over the surface of the galaxy, allowed us to study the light
distribution and the sharp drops at the end of the bar.
We also made radial profiles of the colour, position angle, ellipticity
and shape. We found that there are universal formats for the two
latter, linked to the form and properties of the bar. In particular,
we find that five of our galaxies have profiles such as those
of MH-type $N$-body bars (AM02), that is a high axial
ratio, near-constant within a considerable radial region, followed by
a very steep drop. The corresponding b$_4$ profiles showed the
existence of strong rectangular-like shapes. All these, taken
together, suggest that a considerable amount of angular momentum has
been exchanged within these galaxies,
i.e., that their haloes have resonances that are capable of absorbing
considerable amounts of angular momentum \citep{ath03}. On the other
hand, three of our galaxies have MD-type characteristics, namely an
ellipticity profile with a maximum at low values and no sharp drop, and very low
values of the b$_4$. We believe that the halo of these galaxies has
been able to exchange considerably less angular momentum than in the
previous cases. These first results will
be followed by a thorough comparison with $N$-body bars, to gauge
better the halo properties, the angular momentum exchange and its
effect on bar growth and slow down.

We have also measured several characteristic scale-lengths. The peak
of b$_4$ and of the ellipticity are at a radius well within the
bar. This is likewise true for the end of the ledges on the radial photometric profiles
due to the bar ($L_{{\rm prof}_i}$). The radius at which the bump in the
luminosity profile due to the bar smoothly joins up to the disc profile ($L_{{\rm prof}_f}$)
is closer to the end of the bar, but usually difficult to determine.
We found four scale-lengths whose values roughly
coincide, and which could be used to measure the bar length, namely
$L_{\rm phase}$, $L_{{\rm drop}}$, $L_{\epsilon_{\rm min}}$ and
$L_{{\rm prof}_f}$, although the measurements of the last one are much
more noisy than those of the other three. Furthermore, we found that $L_{{\rm drop}}$ coincides
with the position where the bar reaches its maximum boxyness, strongly arguing
in favour of this scale-length being an accurate measurement of the bar length
in MH-type galaxies, where this drop is pronounced. Nonetheless, it is
important to have more than one such measure for two main
reasons. First, in order to be able to make comparisons and averages
to diminish the effect of uncertainties (AM02). Second, because according to the
characteristics of the galaxy some of these measurements are not possible. We discussed
in length some such cases here. The link of these scale-lengths with
the resonances will be made with the help of $N$-body bars elsewhere.

Simulations predict that MH-like bars should be longer than MD-like bars
(AM02). Our measurements of the bar length
bear this prediction out. Indeed, we find that the median value for the
bar length in the MH sample is 8.3 Kpc, compared to a median bar
length of 4 Kpc for the MD galaxies.

The assumption that the outer part of the bar is vertically thin allowed us to
obtain formulae for the deprojected bar length, ellipticity and position angle. We have
shown that these estimations agree very well with the values obtained from the
deprojected image for all our galaxies where we consider the latter to
be reliable. This confirms a theoretical prediction coming from both
orbital structure theory and from $N$-body simulations, namely that
while the inner part of the bar might be vertically thick the outer
part is vertically thin \citep{ath05b}. Since these
are intrinsic properties of the bar, independent of our viewing angle,
the values obtained from our formulae should be valid even for
cases of highly inclined galaxies, where the deprojected image might not be
reliable.

In a future paper we will apply
the same surface photometry techniques used here to a suitable sample of $N$-body simulations of
barred galaxies, similar to those presented in AM02. We then intend to address
whether or not a distinction between MH and MD bars can in fact be done in real galaxies
as well, judging from a thorough comparison between the surface photometry results in both real
galaxies and simulations, including different techniques of image decomposition
\citep[e.g.][]{des04,lau05,but06}.
If successful, this approach could allow us to obtain information on
the halo component and on the angular momentum exchange within the
galaxy directly from surface photometry and morphology.
We will also include in our comparisons the vertical {\em kinematics} in observed and model
bars, using, e.g. the measurements of \citet{gad05}.

\section*{Acknowledgments}
It is a pleasure to thank an anonymous referee for many useful comments and remarks
and Ra\'ul M\'ujica for his efforts, which made
this work possible.
We also thank ECOS and ANUIES for financing the exchange project M04U01, and
FAPESP for grants 03/07099-0 and 00/06695-0. DAG, EA and AB would
like to thank INAOE for their kind hospitality, both in the Tonanzintla
and the Cananea sites. DAG thanks the CNRS for a 6 month
poste rouge during which the data were analysed and the project started.
DAG is supported by the Deutsche Forschungsgemeinschaft priority program 1177 (``Witnesses of Cosmic
History: Formation and evolution of galaxies, black holes and their environment'') and the Max Planck
Society. CANICA was developed under CONACYT project G28586E (PI: L. Carrasco).
This research has made use of the
NASA/IPAC Extragalactic Database (NED), which is operated by the Jet Propulsion
Laboratory, California Institute of Technology, under contract with the National Aeronautics
and Space Administration. This research has also made use of NASA's Astrophysics Data System
and of the HyperLeda database (http://leda.univ-lyon1.fr/).

\clearpage

\begin{table*}
\centering
\begin{minipage}{166mm}
\caption{Properties of the sample galaxies.}
\label{sample}
\begin{tabular}{@{}llccccccccl@{}}
\hline \hline
Galaxy & Type & M$_{\rm B}$ & m$_{\rm B}$ & $cz$ & Distance & D$_{25}/2$ & Inclination & PA$_{\rm ln}$ &
${\rm PA}_{\rm ln}-{\rm PA}_{\rm bar}$ & AGN \\
\omit & \omit & \omit & \omit & (Km/s) & (Mpc) & (arcmin) & (deg) & (deg) & (deg) & \omit \\
(1) & (2) & (3) & (4) & (5) & (6) & (7) & (8) & (9) & (10) & (11) \\
\hline
NGC 266  & SB(rs)ab    & $-22.0$ & 12.6 & 4770 & 68.1 & 1.5 & 13.7 & 150 & 50 & \dots \\
NGC 357  & SB(r)0/a    & $-20.2$ & 13.1 & 2379 & 34.0 & 1.1 & 44.9 & 20  & 80 & LINER \\
NGC 799  & (R')SB(s)a  & $-20.7$ & 14.1 & 5846 & 83.5 & 1.0 & 34.4 & 100 & 55 & \dots \\
NGC 1211 & (R)SB(r)0/a & $-20.1$ & 13.5 & 3132 & 44.7 & 0.9 & 46.2 & 30  & 60 & \dots \\
NGC 1358 & SAB(r)0/a   & $-20.9$ & 13.2 & 3924 & 56.1 & 1.1 & 53.6 & 15  & 60 & Sey2  \\
NGC 1638 & SAB(rs)0$^o$?    & $-20.5$ & 13.1 & 3209 & 45.8 & 1.1 & 55.9 & 70  & 4  & \dots \\
NGC 7080 & SB(r)b      & $-21.4$ & 13.6 & 4998 & 71.4 & 0.9 & 19.6 & 0   & 75 & \dots \\
NGC 7280 & SAB(r)0$^+$     & $-19.4$ & 13.1 & 1942 & 27.7 & 1.0 & 54.4 & 78  & 12  & AGN   \\
NGC 7743 & (R)SB(s)0$^+$   & $-19.7$ & 12.6 & 1725 & 24.6 & 1.4 & 31.0 & 80  & $-$20 & Sey2  \\
\hline
\end{tabular}
Column (1) identifies the galaxy and
column (2) gives its morphological type according to the RC3 \citep{dev91}. Columns (3) and (4)
show, respectively, the absolute and apparent B-band magnitude, according to the Lyon
Extragalactic Data Archive (LEDA). In column (5) the LEDA radial velocity in Km/s, corrected
for the infall of the Local Group towards Virgo, is displayed, and the galaxy distance in Mpc,
using H$_0=70$ Km s$^{-1}$ Mpc$^{-1}$, appears in column (6). Column (7) shows the radius
of the 25 B-band mag arcsec$^{-2}$ isophote according to LEDA, in arcminutes.
Column (8) gives the inclination angle of the plane of the galaxy to the plane of the sky,
in degrees, as in LEDA, except for NGC 7743 to which this parameter was derived in Sect. 4.4.
The position angle of the line of nodes (from North to East) and the difference between this position
angle and that of the bar, i.e., the parameter $\alpha$ in our analytical treatments, are shown in columns (9) and (10),
respectively, in degrees (see Sect. 6 for details). Finally, column (11) shows the AGN designation as
given in the NASA Extragalactic Database (NED).
\end{minipage}
\end{table*}

\begin{table*}
\centering
\begin{minipage}{103mm}
\caption{Summary of the observations.}
\label{obslog}
\begin{tabular}{@{}llccc@{}}
\hline \hline
Night & Galaxy                    & Photometric? & Error (J) & Error (Ks) \\
22/09 & N7280;N1211Ks             & Yes          & 0.05      & 0.06 \\
23/09 & N7280J;N7743Ks;N1211Ks    & Yes          & 0.11      & 0.14 \\
24/09 & N7743J;N1211J             & Yes          & 0.06      & 0.10 \\
25/09 & N7080Ks;N1358Ks           & No           & 0.06      & 0.12 \\
26/09 & N7080Ks;N1358             & No           & 0.06      & 0.09 \\
27/09 & N7080;N1358J;N1638Ks      & No           & 0.07      & 0.05 \\
28/09 & N7080J;N1638              & No           & 0.13      & 0.15 \\
30/09 & N7080J;N1638J             & Yes          & 0.05      & 0.07 \\
01/10 & N357                      & Yes          & 0.06      & 0.06 \\
02/10 & N266Ks;N357J              & Yes          & 0.01      & 0.11 \\
03/10 & N266J;N799Ks              & Yes          & 0.06      & 0.04 \\
04/10 & N799                      & Yes          & 0.10      & 0.10 \\
\hline
\end{tabular}
For each night we list the galaxies observed with the corresponding band appended to the
name of the galaxy; when no band information is given images in both bands were taken.
Photometric errors are given in magnitudes.
\end{minipage}
\end{table*}

\begin{table*}
\centering
\begin{minipage}{177mm}
\caption{Estimates for bar lengths from direct images in arcsec (left) and Kpc
(right).}
\label{lbardirect}
\begin{tabular}{@{}lccccccc|ccccccc@{}}
\hline \hline
Galaxy & $L_{\epsilon_{\rm max}}$ & $L_{{\rm drop}}$ & $L_{\epsilon_{\rm min}}$ &
$L_{{\rm phase}}$ & $L_{{\rm b_4}}$ & $L_{{\rm prof}_i}$ & $L_{{\rm prof}_f}$ &
$L_{\epsilon_{\rm max}}$ & $L_{{\rm drop}}$ & $L_{\epsilon_{\rm min}}$ &
$L_{{\rm phase}}$ & $L_{{\rm b_4}}$ & $L_{{\rm prof}_i}$ & $L_{{\rm prof}_f}$\\
\omit & \omit & \omit &  \omit &  (arcsec) & \omit & \omit & \omit & \omit & \omit & \omit &  (Kpc) &  \omit & \omit & \omit \\
\hline
N266    & 37.9 & 49.1  & 49.1 & 48.5   & 25.2  & 30.6  & 35.0       & 12.5 & 16.2  & 16.2  & 16.0  & 8.3   & 10.1   & 11.6  \\
N266b   & 63.0 & 66.0  & 67.0 & 67.0   & 53.0  & \dots & \dots      & 20.8 & 21.8  & 22.1  & 22.1  & 17.5  & \dots  & \dots \\
N357    & 21.2 & 24.8  & 27.3 & 26.7   & 17.0  & 18.8  & 23.3       & 3.5  & 4.1   & 4.5   & 4.4   & 2.8   & 3.1    & 3.8   \\
N799    & 14.1 & \dots & 21.0 & 21.0   & 11.1  & \dots & \dots      & 5.7  & \dots & 8.5   & 8.5   & 4.5   & \dots  & \dots \\
N1211   & 24.9 & 28.1  & 31.8 & 48.4   & 17.1  & 18.9  & 30.0       & 5.4  & 6.1   & 6.9   & 10.5  & 3.7   & 4.1    & 6.5   \\
N1211b  & 42.9 & 48.8  & 48.8 & 48.8   & 35.0  & \dots & \dots      & 9.3  & 10.6  & 10.6  & 10.6  & 7.6   & \dots  & \dots \\
N1358   & 15.1 & 23.0  & 24.2 & 23.2   & 11.0  & 12.9  & 15.4       & 4.1  & 6.2   & 6.6   & 6.3   & 3.0   & 3.5    & 4.2   \\
N7080   & 17.9 & 22.0  & 22.0 & 21.4   & 13.9  & 17.9  & 20.1       & 6.2  & 7.6   & 7.6   & 7.4   & 4.8   & 6.2    & 7.0   \\
N7080b  & 28.9 & 32.1  & 35.0 & 26.0   & 26.0  & 23.1  & \dots      & 10.0 & 11.1  & 12.1  & 9.0   & 9.0   & 8.0    & \dots \\
N7280   & 14.2 & 20.9  & 23.1 & 32.1   & \dots & \dots & \dots      & 1.9  & 2.8   & 3.1   & 4.3   & \dots & \dots  & \dots \\
N7743   & 42.9 & 45.4  & 57.1 & 47.9   & 36.1  & 40.3  & \dots      & 5.1  & 5.4   & 6.8   & 5.7   & 4.3   & 4.8    & \dots \\
\hline
\end{tabular}
Bar length estimates for each galaxy: $L_{\epsilon_{\rm max}}$ is the position of the maximum of
the ellipticity profile; $L_{{\rm drop}}$ is the last position before the maximum
change in slope in the ellipticity profile just after $L_{\epsilon_{\rm max}}$;
$L_{\epsilon_{\rm min}}$ is the position of the first ellipticity minimum after
$L_{\epsilon_{\rm max}}$; $L_{{\rm phase}}$ is the position where the position angle
of the isophotes changes by more than $10^{\circ}$ from that of the bar; $L_{{\rm b_4}}$
is the position of the maximum in b$_4$, and, finally, in the last two columns we give the length estimates
from the luminosity profile, $L_{{\rm prof}_i}$ and $L_{{\rm prof}_f}$ (see text for further details).
In some cases no reliable estimate was possible. Estimates for
the second sub-structure in NGC 266, 1211 and NGC 7080 are also given (see text for details).
\end{minipage}
\end{table*}

\begin{table*}
\centering
\begin{minipage}{177mm}
\caption{Estimates for bar lengths from deprojected images in arcsec (left) and Kpc
(right).}
\label{lbardeproj}
\begin{tabular}{@{}lccccccc|ccccccc@{}}
\hline \hline
Galaxy & $L_{\epsilon_{\rm max}}$ & $L_{{\rm drop}}$ & $L_{\epsilon_{\rm min}}$ &
$L_{{\rm phase}}$ & $L_{{\rm b_4}}$ & $L_{{\rm prof}_i}$ & $L_{{\rm prof}_f}$ &
$L_{\epsilon_{\rm max}}$ & $L_{{\rm drop}}$ & $L_{\epsilon_{\rm min}}$ &
$L_{{\rm phase}}$ & $L_{{\rm b_4}}$ & $L_{{\rm prof}_i}$ & $L_{{\rm prof}_f}$\\
\omit & \omit & \omit &  \omit &  (arcsec) & \omit & \omit & \omit & \omit & \omit & \omit &  (Kpc) &  \omit & \omit & \omit \\
\hline
N266    & 41.8  & 49.1  & 49.1  & 49.1   & 23.6  & 30.1   & 35.0        & 13.8 & 16.2  & 16.2  & 16.2  & 7.8   & 10.2   & 11.9   \\
N266b   & 62.1  & 67.0  & 67.9  & 67.9   & 53.0  & \dots  & \dots       & 20.5 & 22.1  & 22.4  & 22.4  & 17.5  & \dots  & \dots  \\
N357    & 31.5  & 35.8  & 38.8  & 50.9   & 23.6  & 23.0   & 38.2        & 5.2  & 5.9   & 6.4   & 8.4   & 3.9   & 3.8    & 6.3    \\
N799    & 17.0  & \dots & 23.0  & (23.0) & 9.9   & \dots  & \dots       & 6.9  & \dots & 9.3   & (9.3) & 4.0   & \dots  & \dots  \\
N1211   & 34.1  & 37.8  & 42.4  & (67.7) & 23.5  & 24.0   & 41.0        & 7.4  & 8.2   & 9.2   & (14.7)& 5.1   & 5.2    & 8.9    \\
N1211b  & 67.7  & \dots & \dots & \dots  & 49.8  & \dots  & \dots       & 14.7 & \dots & \dots & \dots & 10.8  & \dots  & \dots  \\
N1358   & 24.6  & 44.9  & 47.8  & 44.9   & 16.9  & 21.0   & 24.6        & 6.7  & 12.2  & 13.0  & 12.2  & 4.6   & 5.7    & 6.7    \\
N7080   & 16.5  & 22.8  & 22.8  & 23.1   & 14.2  & 17.3   & 22.1        & 6.7  & 7.9   & 7.9   & 8.0   & 4.9   & 6.0    & 7.7    \\
N7080b  & 30.9  & 32.9  & 38.2  & 28.0   & 28.0  & 25.1   & \dots       & 10.7 & 11.4  & 13.2  & 9.7   & 9.7   & 8.7    & \dots  \\
N7280   & 28.4  & 32.8  & 50.0  & 29.9   & \dots & \dots  & \dots       & 3.8  & 4.4   & 6.7   & 4.0   & \dots & \dots  & \dots  \\
N7743   & 44.0 & 47.1 & 57.9 & 48.7   & 37.2  & 41.3      & \dots       & 5.2 & 5.6 & 6.9  & 5.8   & 4.4   & 4.9        & \dots  \\
\hline
\end{tabular}
Same as Table 3 but for the deprojected images. The values in parenthesis for $L_{{\rm phase}}$ come from the
assumption that our photometric measurements reach just this point.
\end{minipage}
\end{table*}

\begin{table*}
\centering
\begin{minipage}{80mm}
\caption{Estimates for bar lengths from direct images normalised by $L_{{\rm phase}}$.}
\label{lbarratiodirect}
\begin{tabular}{@{}lcccccc@{}}
\hline \hline
Galaxy & $L_{\epsilon_{\rm max}}$ & $L_{{\rm drop}}$ & $L_{\epsilon_{\rm min}}$ &
$L_{{\rm b_4}}$ & $L_{{\rm prof}_i}$ & $L_{{\rm prof}_f}$ \\
N266   & 0.78 & 1.01  & 1.01  & 0.52  & 0.63   & 0.72  \\
N357   & 0.80 & 0.93  & 1.02  & 0.64  & 0.70   & 0.87  \\
N799   & 0.67 & \dots & 1.00  & 0.53  & \dots  & \dots \\
N1211  & 0.92 & 1.04  & 1.18  & 0.63  & 0.67   & 1.11  \\
N1358  & 0.65 & 0.99  & 1.04  & 0.48  & 0.56   & 0.66  \\
N7080  & 0.84 & 1.03  & 1.03  & 0.65  & 0.84   & 0.94  \\
N7280  & 0.68 & 1.00  & 1.10  & \dots & \dots  & \dots \\
\hline
\end{tabular}
Same as Table 3 but with all measurements normalised by $L_{{\rm phase}}$.
\end{minipage}
\end{table*}

\begin{table*}
\centering
\begin{minipage}{80mm}
\caption{Estimates for bar lengths from deprojected images normalised by
$L_{{\rm phase}}$.}
\label{lbarratiodeproj}
\begin{tabular}{@{}lcccccc@{}}
\hline \hline
Galaxy & $L_{\epsilon_{\rm max}}$ & $L_{{\rm drop}}$ & $L_{\epsilon_{\rm min}}$ &
$L_{{\rm b_4}}$ & $L_{{\rm prof}_i}$ & $L_{{\rm prof}_f}$\\
N266   & 0.85 & 1.00  & 1.00  & 0.48  & 0.63  & 0.71  \\
N357   & 0.62 & 0.70  & 0.76  & 0.46  & 0.45  & 0.75  \\
N799   & 0.74 & \dots & 1.00  & 0.43  & \dots & \dots \\
N1211  & 0.92 & 1.02  & 1.15  & 0.63  & 0.65  & 1.11  \\
N1358  & 0.55 & 1.00  & 1.07  & 0.38  & 0.47  & 0.55  \\
N7080  & 0.84 & 0.99  & 0.99  & 0.61  & 0.75  & 0.96  \\
N7280  & 0.95 & 1.10  & 1.68  & \dots & \dots & \dots \\
\hline
\end{tabular}
Same as Table 5 but for the deprojected images.
\end{minipage}
\end{table*}

\begin{table*}
\centering
\begin{minipage}{125mm}
\caption{Results from the analytical deprojection of bars.}
\label{lbaranl}
\begin{tabular}{@{}ccccccccc@{}}
\hline \hline
\omit      & N266$^a$  & N357$^a$  & N799$^b$  & N1211$^a$  & N1358$^a$ & N7080$^a$ & N7280$^b$
& N7743$^c$ \\
\hline
\omit&\omit&\omit&\omit& 2D&\omit&\omit&\omit&\omit\\
Bar length & \omit & \omit & \omit & \omit & \omit  & \omit & \omit & \omit \\
(arcsec)   & 50.0  & 34.8  & 24.3  & 38.3  & 36.4   & 23.3  & 24.5  & 33.8  \\
Bar length & \omit & \omit & \omit & \omit & \omit  & \omit & \omit & \omit \\
(Kpc)      & 16.5  & 5.7   & 9.8   & 8.3   & 9.9    & 8.1   & 3.3   & 4.0   \\
Ellipticity& 0.59  & 0.58  & 0.41  & 0.60  & 0.58   & 0.59  & 0.21  & 0.38  \\
${\rm PA}_{\rm ln}-{\rm PA}_{\rm bar}$& \omit & \omit & \omit & \omit & \omit  & \omit & \omit & \omit \\
(deg) & 51  & 84  & 66  & 72  & 77   & 76  & 62  & $-$27 \\
\hline
\omit&\omit&\omit&\omit& 1D&\omit&\omit&\omit&\omit\\
Bar length & \omit & \omit & \omit & \omit & \omit  & \omit & \omit & \omit \\
(arcsec)   & 49.9  & 34.7  & 24.1  & 37.9  & 35.5   & 23.3  & 21.9  & 33.7  \\
Bar length & \omit & \omit & \omit & \omit & \omit  & \omit & \omit & \omit \\
(Kpc)      & 16.5  & 5.7   & 9.7   & 8.2   & 9.6    & 8.0   & 2.9   & 4.0   \\
\hline
\end{tabular}
$^a$ $L_p=$ $L_{{\rm drop}}$\\
$^b$ $L_p=$ $L_{{\rm phase}}$\\
$^c$ $L_p$ is taken as the position of the glitch in the ellipticity profile (see Sect. 6.1).
\end{minipage}
\end{table*}

\clearpage

\appendix
\onecolumn
\section{Analytical Deprojection of an Ellipse}

In the following, we will derive analytical expressions for the true semi-major and semi-minor axes and
position angle of an ellipse seen in projection\footnote{The source code of a
{\sc fortran} program to perform these calculations can be downloaded
at http://www.mpa-garching.mpg.de/$\sim$dimitri/deprojell.f .}.

In the reference frame of an ellipse, i.e., in a
coordinate system $(s,t)$ centred at the ellipse centre and with the axis of the abscissae $s$ aligned with
the ellipse major axis, one can write:

\begin{equation}
\frac{s^2}{a^2}+\frac{t^2}{b^2}=1,
\end{equation}

\noindent where $a$ and $b$ are the ellipse semi-major and semi-minor axes. Considering a coordinate
system $(x,y)$, rotated with respect to the ellipse coordinate system but also centred at the ellipse
centre, one can show that:

\begin{equation}
\begin{array}{l}
s=x\cos\alpha+y\sin\alpha \\
t=y\cos\alpha-x\sin\alpha,
\end{array}
\end{equation}

\noindent where $\alpha$ is the angle between the two coordinate systems, counted counter-clockwise from
$(x,y)$ to $(s,t)$. Substituting Eqs. (A2) into Eq. (A1), it is possible to obtain the equation of the
ellipse in the $(x,y)$ coordinate system in its quadratic form:

\begin{equation}
Ax^2+2Bxy+Cy^2+2Dx+2Fy+G=0,
\end{equation}

\noindent where

\begin{equation}
A=\frac{\cos^2\alpha}{a^2}+\frac{\sin^2\alpha}{b^2}
\end{equation}

\begin{equation}
B=\frac{\cos\alpha\sin\alpha}{a^2}-\frac{\cos\alpha\sin\alpha}{b^2}
\end{equation}

\begin{equation}
C=\frac{\sin^2\alpha}{a^2}+\frac{\cos^2\alpha}{b^2}
\end{equation}

\noindent and $D=F=0$ and $G=-1$.

Now, consider that the ellipse and its reference frame $(s,t)$ are rotated about the $x$-axis by
an angle $i$. The projection of the inclined ellipse onto the plane given by $(x,y)$ gives another
ellipse, whose equation is identical to Eq. (A3), except that $y$ is replaced by $y\cos i$.
It is straightforward to show that the equation of the deprojected ellipse is:

\begin{equation}
A^\prime x^2+2B^\prime xy+C^\prime y^2+2D^\prime x+2F^\prime y+G^\prime=0,
\end{equation}

\noindent where

\begin{equation}
B^\prime=B\cos i
\end{equation}

\begin{equation}
C^\prime=C\cos^2i,
\end{equation}

\noindent and $A^\prime=A$, $D^\prime=F^\prime=0$ and $G^\prime=G=-1$.

The semi-major and semi-minor axes of an ellipse, as well as its position angle,
can be directly obtained from its quadratic equation. For the deprojected ellipse,

\begin{equation}
s1=\left( \frac{2(A^\prime F^\prime{}^2+C^\prime D^\prime{}^2+G^\prime B^\prime{}^2-2B^\prime D^\prime
F^\prime-A^\prime C^\prime G^\prime)}{(B^\prime{}^2-A^\prime C^\prime)\left[ (C^\prime-A^\prime)
\sqrt{1+\frac{4B^\prime{}^2}{(A^\prime-C^\prime)^2}}-(C^\prime+A^\prime) \right]}  \right)^\frac{1}{2}
\end{equation}

\noindent and

\begin{equation}
s2=\left( \frac{2(A^\prime F^\prime{}^2+C^\prime D^\prime{}^2+G^\prime B^\prime{}^2-2B^\prime D^\prime
F^\prime-A^\prime C^\prime G^\prime)}{(B^\prime{}^2-A^\prime C^\prime)\left[ (A^\prime-C^\prime)
\sqrt{1+\frac{4B^\prime{}^2}{(A^\prime-C^\prime)^2}}-(C^\prime+A^\prime) \right]}  \right)^\frac{1}{2},
\end{equation}

\noindent and the semi-major and semi-minor axes are given, respectively, by $a^\prime=\max(s1,s2)$ and
$b^\prime=\min(s1,s2)$. Thus, the ellipticity is $1-b^\prime/a^\prime$.

The position angle of the deprojected ellipse is given by:

\begin{equation}
\theta=-\frac{1}{2}\cot^{-1}\left(\frac{C^\prime-A^\prime}{2B^\prime}\right)
\end{equation}

\noindent and counted counter-clockwise from the axis in $(x,y)$ which is closer to the ellipse
major axis.

Note that Eqs. (A10) and (A11) have singularities when $i=0$ and $\alpha=\pm n\pi/4$ ($n$ being a positive
integer). A singularity also appears in Eq. (A12) when $\alpha=0,\pm n\pi/2$. If $i=\pi/2$ the three equations
diverge.
For the practical purpose of obtaining deprojected measurements of the properties of galactic bars
from ellipse fits, $\alpha$ is the angle between the bar and the line of nodes, and $i$ is the galaxy
inclination angle.

\bsp

\label{lastpage}

\end{document}